\documentclass[a4paper,11pt]{article}
\usepackage{jheppub}

\usepackage{amssymb}
\usepackage{amsmath}
\usepackage{graphicx}
\usepackage{subcaption}
\usepackage[colorlinks=true]{hyperref}
\usepackage{appendix}
\usepackage{amsbsy}
\usepackage{slashed}
\usepackage{tikz}
\usepackage{tikz-feynman}

\def\e{\epsilon}
\newcommand{\code}[1]{\textsc{#1}}

\preprint{{\raggedleft
		ZU-TH 21/23, TUM-HEP 1459/23\\
}}

\title{\boldmath Two-loop helicity amplitudes for $V+$jet production including axial vector couplings to higher orders in $\epsilon$}

\author[a]{Thomas Gehrmann,}
\author[a]{Petr Jakub\v{c}\'{i}k,}
\author[b]{Cesare Carlo Mella,}
\author[b]{Nikolaos Syrrakos,}
\author[b]{Lorenzo Tancredi}

\affiliation[a]{Physik-Institut, Universit\"{a}t Zurich, 
            Winterthurerstrasse 190,
           	CH-8057 Z\"{u}rich,
            Switzerland}
\affiliation[b]{Technical University of Munich, TUM School of Natural Sciences, Physics Department, James-Franck-Straße 1, 85748 Garching, Germany}
            
\emailAdd{thomas.gehrmann@physik.uzh.ch}            
\emailAdd{petr.jakubcik@physik.uzh.ch}
\emailAdd{cesarecarlo.mella@tum.de}
\emailAdd{nikolaos.syrrakos@tum.de}
\emailAdd{lorenzo.tancredi@tum.de}

\abstract{ 
We compute the two-loop Quantum Chromodynamics (QCD) corrections to 
all partonic channels relevant for the production of 
an electroweak boson $V=Z,W^\pm,\gamma^*$ and a jet at 
hadron colliders. 
We consider
the decay of a vector boson $V$ to three partons
$ V \to  q\bar{q}g$, $ V \to ggg$ with a 
vector and axial vector coupling in both channels,
including singlet and non-singlet contributions.
For the quark channel,
we use a recent tensor decomposition and
extend the calculation to $\mathcal{O}(\epsilon^2)$.
For the gluonic channel, we define a new
tensor decomposition which allows us to compute the vector and the axial vector amplitudes at once and to perform
the computation of the amplitudes to $\mathcal{O}(\epsilon^2)$. 
We provide finite remainders of the helicity amplitudes analytically continued to all relevant scattering regions 
$q\bar{q} \to V g$, $q g \to V q$ and $gg \to V g$. The axial vector contribution to the gluon-induced
channel completes the set of two-loop amplitudes for this process, while the extension to $\mathcal{O}(\epsilon^2)$ represents the first step in the calculation of next-to-next-to-next-to-leading-order (N$^3$LO) QCD corrections to $Z$+jet production at hadron colliders.}

\begin{document}
       
\maketitle

\section{Introduction}\label{sec:intro}
The production of an electroweak vector boson $V$ in association with a jet at the Large Hadron Collider (LHC) provides unique
opportunities to test Quantum Chromodynamics (QCD) 
and the Standard Model of particle physics (SM) to unprecedented precision.
Especially in the case of $V=Z,\gamma^*$, the vector bosons are 
easy to reconstruct through the clear signatures of their leptonic decays
in detectors. Standard observables, such as the 
transverse momentum of the lepton pair $p_T^{\ell \ell}$ into which they decay and their invariant-mass spectrum,
can therefore be measured to precisions well below the percent level~\cite{ATLAS:2014alx,CMS:2015hyl,CMS:2011wyd,ATLAS:2015iiu,LHCb:2015okr}.
The large amount of data
available also makes it possible to measure more differential observables to very high precision,
including angular distributions of the leptonic decay products of the $V$ bosons~\cite{CMS:2011utm,ATLAS:2015ihy,LHCb:2015jyu}.
These provide a direct way to investigate the polarisation of the gauge bosons,
which in turn depends on the details of the dynamics of its production mechanisms.

On the theoretical side, a multitude of observables which involve the production of an electroweak boson at the LHC 
can be described  
with uncertainties below the $5\%$ level.
In particular, inclusive and fiducial cross-sections have recently been computed to N$^3$LO in QCD~\cite{Duhr:2020seh,Duhr:2020sdp,Duhr:2021vwj,Chen:2021vtu,Chen:2022cgv}.
Moreover, the production of $V$ bosons in association with resolved QCD radiation can be modelled to next-to-next-to-leading order (NNLO), which includes fully differential predictions for their transverse momentum and the angular coefficients~\cite{Boughezal:2015dva,Gehrmann-DeRidder:2015wbt,Boughezal:2016isb,Gehrmann-DeRidder:2017mvr,Gauld:2017tww,Pellen:2022fom}. 

For most of these observables, matching the astonishing experimental precision of the most recent LHC measurements 
with perturbative calculations
remains out of reach due
to unsurmountable difficulties in controlling non-perturbative effects in QCD below the $\mathcal{O}(1\%)$ level.
Nevertheless, theory predictions can be improved by handling different sources of uncertainties at very high precision. In particular, 
for calculations which are 
differential in QCD radiation, a complete account of 
N$^3$LO QCD corrections to $V$+jet production is necessary.
One of the ingredients which enters this calculation are the finite remainders of the three loop
QCD amplitudes for the production of a $V$ and a parton in partonic collisions. 
While these remain an outstanding task, important progress has been already achieved with the calculation
of all relevant planar~\cite{DiVita:2014pza,Canko:2021xmn} and some non-planar~\cite{Henn:2023vbd} families of the relevant three-loop Feynman integrals.
Importantly, the evaluation of the three-loop virtual finite remainder also requires the corresponding virtual amplitudes evaluated at
lower loop orders but at higher orders in the dimensional regulators $\epsilon$. The two-loop master integrals and amplitudes were previously computed~\cite{Gehrmann:2000zt,Gehrmann:2001ck,Garland:2002ak,Gehrmann:2022vuk} only through to finite terms in $\epsilon$.
In this paper, we address this problem, evaluating the building blocks of two-loop amplitudes for $q\bar{q} \to V g$, $q g \to V q$ and $gg \to Vg$ 
up to $\mathcal{O}(\epsilon^2)$, using the two-loop master integrals
recently recomputed to this order in~\cite{Gehrmann:2023etk}.

Additionally, we take the opportunity to provide the last outstanding contribution to the production of a $V$ boson and a jet at this loop order in QCD perturbation theory,
namely the two-loop corrections to $gg \to V g$, where the vector boson couples to a virtual quark loop through an axial vector current.
These amplitudes vanish for $V = W^\pm$
due to charge conservation and, moreover, they are expected to cancel also for $V=Z$ when summing over degenerate isospin doublets, such that only the contribution
of (massless) bottom quarks has to be taken into account. These axial vector corrections are expected to be negligible for inclusive observables. Nevertheless, their impact might become sizeable if sufficiently differential observables are considered,  in particular when compared to the N$^3$LO corrections which will become available in the future.

The corresponding axial vector amplitudes for the quark-antiquark channel $V \to q\bar{q}g$ have recently been computed in~\cite{Gehrmann:2022vuk}. In this case, one could separate the amplitude with axial vector coupling into a flavour non-singlet 
and pure-singlet contribution. The former involves all diagrams where the vector boson couples to the external fermion line, while in the latter, the vector boson couples to a loop of virtual quarks.
In this channel, the pure-singlet contributions are  well-defined and anomaly-free at the two-loop order only if
loops of top quarks are also taken into account.

In the gluonic channel considered here, only a single axial vector current enters the amplitudes: the flavour singlet. The absence of external fermions does not allow for a decomposition into flavour non-singlet and pure-singlet. We therefore treat the amplitude as a whole and refrain from denoting it explicitly as the singlet.

Contrary to the quark channel, top-induced contributions are not required up to the two-loop order in the purely gluonic amplitudes, and hence we do not consider them here.

In summary, we provide the tree-level helicity amplitudes for $Vggg$ and $Vgq\bar{q}$, and the finite remainders of the one- and two-loop amplitudes, with vector and axial vector couplings, after renormalisation and IR subtraction, and continued to all production regions. In view of the three-loop calculation, we also give the necessary building blocks extended to transcendental weight six, namely the renormalised helicity amplitudes in the decay region with Minkowski kinematics for both processes and types of coupling.

The rest of this paper is organised as follows. In section~\ref{sec:kinematics}, we define the relevant processes and their kinematics. We start by considering the scattering amplitudes in decay kinematics, where a generic vector boson
$V$ decays to three QCD partons. In section~\ref{sec:tensor}, we then
introduce a new tensor decomposition which aids us in computing the QCD corrections to the amplitudes coupling to a vector and axial vector current. The details of the calculation of the corresponding form factors are described in section~\ref{sec:formfactors}, including the use of the Larin scheme~\cite{Larin:1993tq} for dealing with $\gamma_5$ in dimensional regularisation. UV renormalisation and the structure of IR poles are
explained in section~\ref{sec:UVandIR}. We then move to computing the helicity amplitudes using spinor-helicity formalism in section~\ref{sec:helicity},
including a short description of the procedure required for their analytic continuation to all relevant regions for scattering kinematics. The validation of our results against literature and other consistency checks are described in section~\ref{section:results}.
Finally we draw our conclusions in section~\ref{sec:conclusions}.

\section{Notation and kinematics}\label{sec:kinematics}
We start by considering the decay of a vector boson $V$ into three partons. The two relevant channels are the decay of the $V$ into a quark, anti-quark and a gluon
\begin{equation}
V(p_4) \rightarrow q(p_1) + \bar{q}(p_2) + g(p_3)\,,
\end{equation}
and alternatively into three gluons
\begin{equation}
V(p_4) \rightarrow g_1(p_1) + g_2(p_2) + g_3(p_3)\,.
\end{equation}
Amplitudes for the production processes are then obtained through 
analytic continuation of the decay kinematics~\cite{Gehrmann:2002zr}, see Section \ref{sec:continuations} and~\cite{Gehrmann:2023etk} for more details.

The Mandelstam invariants are defined as
\begin{equation}
s_{12} = (p_1+p_2)^2\,, \quad \quad  s_{13} = (p_1+p_3)^2\,, \quad \quad s_{23} = (p_2+p_3)^2
\end{equation}
and satisfy the conservation equation
\begin{equation}
s_{12} + s_{13} + s_{23} = q^2\,,
\label{eqn:mandelstamsum}
\end{equation}
with $q^2=p_4^2$ being the momentum-squared of the vector boson. We prefer to work with the following dimensionless ratios
\begin{equation}
    x = \frac{s_{12}}{q^2}\,, \quad \quad  y = \frac{s_{13}}{q^2}\,, \quad \quad z = \frac{s_{23}}{q^2}\,, 
    \label{eqn:xyz}
\end{equation}
such that \eqref{eqn:mandelstamsum} implies the relation
\begin{equation}
x+y+z=1\,.
\label{eqn:xyzsum}
\end{equation}
In the decay kinematic region, all these invariants are non-negative. This, together with \eqref{eqn:xyz}, defines the corresponding kinematic region
\begin{equation}
 z \geq 0\,, \quad \quad  0 \leq  y  \leq 1 - z\,, \quad  \quad x = 1 - y -z\,. 
\end{equation}

\section{Tensor decomposition including axial vector terms}\label{sec:tensor}
A multi-loop amplitude can be decomposed into a basis of Lorentz tensor structures, which are subsequently used to construct the required helicity amplitudes. For the process ${Vgq\bar{q}}$ we refer to section 2 of \cite{Gehrmann:2022vuk}, where the relevant tensor basis was derived. For the process $Vggg$, we derive a new basis compared to \cite{Gehrmann:2013vga}, using the construction described in \cite{Peraro:2019cjj, Peraro:2020sfm}. In this framework, one can avoid calculating evanescent tensor structures in $d=4$, and the number of tensor structures matches the number of independent helicity amplitudes. Given two polarisations for each gluon and three for the massive vector boson, there are $2\times 2 \times2 \times 3 = 24$ such helicity configurations. In the absence of parity-breaking terms, only 12 of these helicity amplitudes would be independent, related to the other half by a parity transformation. For the Standard Model $Z$ boson, the symmetry is broken by the presence of a vector-axial coupling, so we expect all 24 structures to be independent.

The construction is performed in the 't Hooft-Veltman dimensional regularization scheme~\cite{tHooft:1972tcz}, which implies 4-dimensional external states. It was shown in~\cite{Peraro:2019cjj, Peraro:2020sfm} that the evanescent tensor structures do not contribute to the helicity amplitudes in this regularization scheme. The amplitude can be written as
\begin{align}
\mathcal{M}_{\omega} &= -i\sqrt{4\pi\bar{\alpha}_s}\, C_{\omega}^{a_1 a_2 a_3} A^{\omega}_{\mu_1 \mu_2 \mu_3 \mu} \epsilon_1^{\mu_1} \epsilon_2^{\mu_2}  \epsilon_3^{\mu_3}\epsilon_4^{\mu}\,,
\end{align}
where $\omega = \{v,a\}$ denote the vector and axial vector contributions, $\mu$ is the index corresponding to the massive vector boson $V$, $a_i$ is the colour index of gluon $i$, $\bar{\alpha}_s$ the unrenormalised strong coupling and $C_{v}^{a_1 a_2 a_3}=d^{a_1 a_2 a_3}$ and $C_{a}^{a_1 a_2 a_3}=f^{a_1 a_2 a_3}$ are the symmetric and anti-symmetric tensors in $SU(3)$. In $d=4$ dimensions, we build a basis of vectors considering the three independent external momenta $\{p_1,p_2,p_3\}$ and the axial vector $v_A$, defined by means of the totally antisymmetric Levi-Civita symbol as
\begin{align}
    v_A^\mu\, \equiv \epsilon^{\nu\rho\sigma\mu}p_{1\nu}p_{2\mu}p_{3\sigma} = \epsilon^{p_1 p_2 p_3 \mu}\,.
    \label{eqn:v_A}
\end{align}
By construction, $v_A$ is orthogonal to the three momenta used to define it, $p_i \cdot v_A = 0$, $i=1,2,3$. Following the FORM~\cite{Vermaseren:2000nd} convention, let us define the Levi-Civita symbol $\epsilon^{\nu\rho\sigma\mu}$ with
\begin{align}
    \epsilon^{0123}& = -\epsilon_{0123} = -i\,, \qquad
    \epsilon^{\nu\rho\sigma\mu}\epsilon_{\nu\rho\sigma\mu} = 24 + \mathcal{O}(d-4)\,.
\end{align}
Defining $q_j = p_j$ for $j=1,2,3$ and $q_4 = v_A$,
the tensorial amplitude can  be decomposed as
\begin{align}
A^{\omega}_{\mu_1 \mu_2 \mu_3 \mu}  = \sum_{i_1 i_2,i_3 i_4} \mathcal{F}^{\omega}_{i_1 i_2 i_3 i_4}q_{i_1}^{\mu_1}q_{i_2}^{\mu_2}q_{i_3}^{\mu_3}q_{i_4}^{\mu}\,,
\end{align}
with $\mathcal{F}^\omega_{i_1 i_2,i_3 i_4}$ scalar form factors.
Crucially, because we are in the 
't Hooft-Veltman scheme, indices not contracted with external momenta will always be taken as $d$-dimensional,
\begin{equation}
g^{\mu \nu} g_{\mu \nu} = d\,, \quad v_A \cdot v_A = 
\frac{d-3}{4} s_{12} s_{13} s_{23}\,. \label{eq:ddimcontr}
\end{equation}

Restricting the possible combinations using gluon transversality and Lorenz gauge for the massive vector boson, $\epsilon_i \cdot p_i = 0\,$, together with the cyclic gauge choice for gluons, $\epsilon_1 \cdot p_2 = \epsilon_2 \cdot p_3 = \epsilon_3 \cdot p_1 = 0$, we are left with the expected 24 tensor structures:

\begin{align}
A^{v}_{\mu_1 \mu_2 \mu_3 \mu}  &= 
+\Tilde{F_1} \, p_3^{\mu_1}p_1^{\mu_2}p_2^{\mu_3}p_1^{\mu} + \Tilde{F_2} \, p_3^{\mu_1}p_1^{\mu_2}p_2^{\mu_3}p_2^{\mu} + \Tilde{F_3} \, p_3^{\mu_1}p_1^{\mu_2}p_2^{\mu_3} v_A^{\mu} \nonumber \\
&\phantom{ =\,\,}+ 
\Tilde{F_4} \, p_3^{\mu_1}p_1^{\mu_2}v_A^{\mu_3}p_1^{\mu} + \Tilde{F_5}\, p_3^{\mu_1}p_1^{\mu_2}v_A^{\mu_3}p_2^{\mu} + \Tilde{F_6} \, p_3^{\mu_1}p_1^{\mu_2}v_A^{\mu_3} v_A^{\mu} \nonumber \\
&\phantom{ =\,\,}+  
\Tilde{F_7} \, p_3^{\mu_1}v_A^{\mu_2}p_2^{\mu_3}p_1^{\mu} + \Tilde{F_8} \, p_3^{\mu_1}v_A^{\mu_2}p_2^{\mu_3}p_2^{\mu} + \Tilde{F_9} \, p_3^{\mu_1}v_A^{\mu_2}p_2^{\mu_3} v_A^{\mu} \nonumber \\
&\phantom{ =\,\,}+  
\Tilde{F}_{10} \, p_3^{\mu_1}v_A^{\mu_2}v_A^{\mu_3}p_1^{\mu} + \Tilde{F}_{11} \, p_3^{\mu_1}v_A^{\mu_2}v_A^{\mu_3}p_2^{\mu} + \Tilde{F}_{12} \, p_3^{\mu_1}v_A^{\mu_2}v_A^{\mu_3} v_A^{\mu}\,, \nonumber \\
A^{a}_{\mu_1 \mu_2 \mu_3 \mu}  &= 
+\Tilde{F}_{13} \, v_A^{\mu_1}p_1^{\mu_2}p_2^{\mu_3}p_1^{\mu} + \Tilde{F}_{14} \, v_A^{\mu_1}p_1^{\mu_2}p_2^{\mu_3}p_2^{\mu} + \Tilde{F}_{15} \, v_A^{\mu_1}p_1^{\mu_2}p_2^{\mu_3} v_A^{\mu} \nonumber \\
&\phantom{ =\,\,}+  
\Tilde{F}_{16}\, v_A^{\mu_1}p_1^{\mu_2}v_A^{\mu_3}p_1^{\mu} + \Tilde{F}_{17} \, v_A^{\mu_1}p_1^{\mu_2}v_A^{\mu_3}p_2^{\mu} + \Tilde{F}_{18} \, v_A^{\mu_1}p_1^{\mu_2}v_A^{\mu_3} v_A^{\mu} \nonumber \\
&\phantom{ =\,\,}+  
\Tilde{F}_{19} \, v_A^{\mu_1}v_A^{\mu_2}p_2^{\mu_3}p_1^{\mu} + \Tilde{F}_{20} \, v_A^{\mu_1}v_A^{\mu_2}p_2^{\mu_3}p_2^{\mu} + \Tilde{F}_{21} \, v_A^{\mu_1}v_A^{\mu_2}p_2^{\mu_3} v_A^{\mu} \nonumber \\
&\phantom{ =\,\,}+  
\Tilde{F}_{22} \, v_A^{\mu_1}v_A^{\mu_2}v_A^{\mu_3}p_1^{\mu} + \Tilde{F}_{23} \, v_A^{\mu_1}v_A^{\mu_2}v_A^{\mu_3}p_2^{\mu} + \Tilde{F}_{24} \, v_A^{\mu_1}v_A^{\mu_2}v_A^{\mu_3} v_A^{\mu} \,.
\end{align}

Notice that for external gluons we chose an axial gauge whereas for the massive boson the Lorenz gauge is used. Accordingly, polarisation sums read
\begin{align}
\sum_{pol} \epsilon_i^{\mu*} \epsilon_i^\nu = 
- g^{\mu \nu} + \frac{p_i^\mu p_{i+1}^\nu + p_i^\nu p_{i+1}^\mu }{p_{i} \cdot p_{i+1}}\,,
\qquad
\sum_{pol} \epsilon_4^{\mu*} \epsilon_4^\nu = 
- g^{\mu \nu} + \frac{p_4^\mu  p_4^\nu  }{q^2}\,,
\label{eq:polsums}
\end{align}
cyclically for the indices $i\in\{1,2,3\}$ of the gluons, and for the vector boson, respectively. 

Whenever multiple parity-odd vectors $v_A^\mu$ are present, the computational complexity increases and an ambiguity arises in the order of contractions between pairs of Levi-Civita tensors when the corresponding projectors are applied to the amplitude. 
This issue can be easily resolved by substituting  all terms where an even number of $v_A^\mu$ vectors occur with the corresponding symmetric combinations of $g^{\mu \nu}$ tensors,
\begin{align}
v_A^{\mu_1}v_A^{ \mu_2 } &\sim g^{\mu_1 \mu_2}\,, \nonumber \\
v_A^{\mu_1}v_A^{ \mu_2 } v_A^{ \mu_3 } &\sim g^{\mu_1 \mu_2} v_A^{\mu_3} +  g^{\mu_1 \mu_3} v_A^{\mu_2} +  g^{\mu_2 \mu_3} v_A^{\mu_1} \equiv \mathcal{T}_O^{\mu_1 \mu_2 \mu_3}\,,
\nonumber \\
v_A^{\mu_1}v_A^{ \mu_2 } v_A^{ \mu_3 }v_A^{ \mu } &\sim g^{\mu_1 \mu_2} g^{\mu_3 \mu}+  g^{\mu_1 \mu_3} g^{\mu_2 \mu} +  g^{\mu_2 \mu_3} g^{\mu_1 \mu} \equiv \mathcal{T}_E^{\mu_1 \mu_2 \mu_3 \mu}\,,
\end{align}
where we used the fact that the tensors on the left-hand side are symmetric under the exchange of any pair of indices. As explained in more detail in~\cite{Peraro:2020sfm,Gehrmann:2022vuk}, 
the new basis of tensors spans the same space.

After a reordering into parity-even tensor structures $T_j^{E,\mu\nu}$ with scalar form factors $F_1,...,F_{12}$ and parity-odd tensor structures $T_j^{O,\mu\nu}$ with form factors $G_1,...,G_{12}$, the tensor decomposition reads

{\allowdisplaybreaks
\begin{align}
A^{v}_{\mu_1 \mu_2 \mu_3 \mu} &= \sum_{i=1}^{12} F_i \,T_i^{E,\mu_1 \mu_2 \mu_3 \mu} \nonumber \\
&= 
+ F_1 \, p_3^{\mu_1}p_1^{\mu_2}p_2^{\mu_3}p_1^{\mu} + F_2 \, p_3^{\mu_1}p_1^{\mu_2}p_2^{\mu_3}p_2^{\mu} + F_3 \, p_3^{\mu_1}p_1^{\mu_2}g^{\mu_3 \mu} \nonumber \\
&\phantom{ =\,\,}+  F_4 \, p_3^{\mu_1}p_2^{\mu_3} g^{\mu_2 \mu}  + F_{5} \, p_3^{\mu_1}p_1^{\mu} g^{\mu_2 \mu_3} + F_{6} \, p_3^{\mu_1}p_2^{\mu} g^{\mu_2 \mu_3}  \nonumber \\
&\phantom{ =\,\,}+  F_{7} \, p_1^{\mu_2}p_2^{\mu_3}  g^{\mu_1 \mu} + F_{8} \,p_1^{\mu_2}p_1^{\mu}  g^{\mu_1 \mu_3} + F_{9} \, p_1^{\mu_2}p_2^{\mu} g^{\mu_1 \mu_3} \nonumber \\
&\phantom{ =\,\,}+  F_{10} \, p_2^{\mu_3}p_1^{\mu} g^{\mu_1 \mu_2}+ F_{11} \, p_2^{\mu_3}p_2^{\mu}g^{\mu_1 \mu_2} + F_{12} \,\mathcal{T}_E^{\mu_1 \mu_2 \mu_3 \mu}, \nonumber \\
A^{a}_{\mu_1 \mu_2 \mu_3 \mu} &=\sum_{i=1}^{12} G_i \,T_i^{O,\mu_1 \mu_2 \mu_3 \mu} \nonumber \\
&= +  G_1 \, p_3^{\mu_1}p_1^{\mu_2}p_2^{\mu_3} v_A^{\mu} + G_2 \, p_3^{\mu_1}p_1^{\mu_2}v_A^{\mu_3}p_1^{\mu} + G_3 \, p_3^{\mu_1}p_1^{\mu_2}v_A^{\mu_3}p_2^{\mu}  \nonumber \\
&\phantom{ =\,\,}+  G_4 \, p_3^{\mu_1}v_A^{\mu_2}p_2^{\mu_3}p_1^{\mu}+ G_5 \, p_3^{\mu_1}v_A^{\mu_2}p_2^{\mu_3}p_2^{\mu} + G_6 \, p_3^{\mu_1} \mathcal{T}_O^{\mu_2 \mu_3 \mu} \nonumber \\
&\phantom{ =\,\,}+ G_7 \, v_A^{\mu_1}p_1^{\mu_2}p_2^{\mu_3}p_1^{\mu} + G_8 \, v_A^{\mu_1}p_1^{\mu_2}p_2^{\mu_3}p_2^{\mu}+ G_9 \, p_1^{\mu_2} \mathcal{T}_O^{\mu_1 \mu_3 \mu} \nonumber \\
&\phantom{ =\,\,}+  G_{10} \,p_2^{\mu_3} \mathcal{T}_O^{\mu_1 \mu_2 \mu} + 
G_{11} \, p_1^{\mu} \mathcal{T}_O^{\mu_1 \mu_2 \mu_3}+ G_{12} \, p_2^{\mu} \mathcal{T}_O^{\mu_1 \mu_2 \mu_3} \,. \label{eq:tensordecomposition}
\end{align}
}

We can define a scalar product on the vector space of the newly defined tensors
\begin{align}
    T_i^{P_1\dagger} \cdot T_j^{P_2} = 
    T_i^{{P_1,{\mu_1} {\mu_2} {\mu_3} {\mu_4}}\dagger} \kappa_{\mu_1 \nu_1; \mu_2 \nu_2; \mu_3 \nu_3; \mu_4 \nu_4}
    T_j^{P_2,{\nu_1} {\nu_2} {\nu_3} {\nu_4}}\,, \label{eq:scalprod}
\end{align}
where $P_i = \{E,O\}$ and the metric is composed of the polarization sums in \eqref{eq:polsums},
\begin{equation}
    \kappa^{\mu_1 \nu_1; \mu_2 \nu_2; \mu_3 \nu_3; \mu_4 \nu_4} =
  \left( - g^{{\mu_4} {\nu_4}} + \frac{p_4^{\mu_4}  p_4^{\nu_4}  }{q^2} \right)
    \prod_{i=1}^3\left( - g^{{\mu_i} {\nu_i}} 
    + \frac{p_{i}^{\mu_i} p_{i+1}^{\nu_i} + p_{i}^{\nu_i} p_{i+1}^{\mu_i} }{p_{i} \cdot p_{i+1}}\right)\,.
\end{equation}
In this metric, the odd and even tensor structures are orthogonal. In fact, in the product of an odd element with an even one, the axial vector can only end up contracted with one of the external momenta which defines it in~\eqref{eqn:v_A} and the result vanishes due to antisymmetry of the Levi-Civita symbol. Therefore, we can define two independent sets of projector operators $\mathcal{P}_{i}^{P}$, which are vectors in the dual space, satisfying
\begin{align}
   \mathcal{P}_i^{P} \cdot T_j^{P} = \delta_{ij} \quad \mbox{for} \quad
   P=\{E,O\}\,.
\end{align}
The projectors can then be expanded in the corresponding dual basis
\begin{align}
    \mathcal{P}_i^{P} = \sum_{j=1}^{12} c_j^{(i),P} T_j^{P\dagger}\,, \label{eqn:proj}
\end{align}
where the coefficients $c_j^{(i),P}$ can be read off from the inverse of the matrix
$(M^{P})_{nm} = T_n^{P\dagger} \cdot T_m^{P} $,
\begin{equation}
    c_j^{(i),P} = \left( M^{P} \right)^{-1}_{ij}\,.
\end{equation}
The algebraic expressions for the projectors are provided in the ancillary files.

\section{Calculation of the form factors}\label{sec:formfactors}
First we computed the unrenormalised tree, one-loop and two-loop contributions to the amplitude through a Feynman diagrams expansion. The terms in  the expansion in the bare strong coupling constant $\bar{\alpha}_s$ can be extracted by acting with projectors directly on this representation, expanding
\begin{align}
\bar{F}_i &= \bar{F}_i^{(0)} + \left( \frac{\bar{\alpha}_s}{2 \pi}\right) \bar{F}_i^{(1)} 
     + \left( \frac{\bar{\alpha}_s}{2 \pi}\right)^2 \bar{F}_i^{(2)} +\mathcal{O}(\bar{\alpha}_s^3) \,,    \\
\bar{G}_i &= \bar{G}_i^{(0)} + \left( \frac{\bar{\alpha}_s}{2 \pi}\right) \bar{G}_i^{(1)} 
     + \left( \frac{\bar{\alpha}_s}{2 \pi}\right)^2 \bar{G}_i^{(2)}  + \mathcal{O}(\bar{\alpha}_s^3)\,.      \label{eq:Funren}
    \end{align}
The relevant diagrams are first generated using \code{QGRAF}~\cite{Nogueira:1991ex}. Every manipulation including Feynman rules, evaluating the Dirac and Lorentz algebra, and the application of projectors were performed in \code{FORM}~\cite{Vermaseren:2000nd}. In order to deal with the axial vector interaction, a consistent treatment of $\gamma_5$ requires the choice of a scheme. We employ the Larin scheme~\cite{Larin:1993tq}, which amounts to substituting the axial vector current with its anti-symmetric version
\begin{align}
\gamma^\mu \gamma^5 \rightarrow \frac{1}{2} (\gamma^\mu \gamma^5 - \gamma^5 \gamma^\mu) = \frac{1}{6} \epsilon^{\mu \nu \rho \sigma} \gamma_\nu \gamma_\rho \gamma_\sigma\,,
\label{eqn:Larin-scheme}
\end{align}
where we changed the overall normalisation to be consistent with our convention for the Levi-Civita symbol. 
Contrary to the orginal definition by 't Hooft and Veltman, no explicit dimensional splitting is required in this scheme. This advantage comes at the price of having to explicitly perform an additional renormalisation of the axial vector current in order to restore the axial Ward identities in dimensional regularization. We stress that in this scheme it is particularly easy to apply the parity-odd projectors to the Feynman diagrams without ever having to manipulate $\gamma_5$. In fact, due to the parity invariance of the form factors and due to the block-diagonal form of the projectors, the parity-even(odd) projectors only need to be applied to the vector(axial vector) part of the Feynman diagrams. In this way, each form factor contains always either zero or two occurences of the Levi-Civita tensor, which in turn can be contracted away in terms of the metric tensor. 

\begin{table}
  \centering
  \begin{tabular}{c|c}
    \multicolumn{1}{c}{Family PL: $\{P_i\}$} & \multicolumn{1}{c}{Family NPL: $\{N_i\}$} \\
    \hline
    \hline
    $k_1$      &   $k_1$       \\
    $k_2$      &   $k_2$       \\
    $k_1 - k_2$   &  $k_1 - p_1$       \\
    $k_1 - p_1$  &    $k_2-p_1$        \\ 
    $k_2 - p_1$  &    $k_1+p_3$       \\
    $k_1-p_1-p_2$ & $k_2+p_3$ \\
    $k_2-p_1-p_2$ & $k_1+p_2+p_3$\\
    $k_1-p_1-p_2-p_3$ & $k_1-k_2$ \\
    $k_2-p_1-p_2-p_3$ & $k_1-k_2+p_2$ \\
  \end{tabular}
  \caption{The two integral families used to map all two-loop Feynman integrals.}\label{tab:auxtopo}
\end{table}

Once the projectors have been applied, all Feynman diagrams can be expressed in terms of scalar integrals, which we classify in the planar and non-planar integral families defined in~\cite{Gehrmann:2023etk}. We report the definition of the families in table~\ref{tab:auxtopo} for convenience. We match all scalar integrals to one of the two families and their kinematic crossings using
\code{Reduze2}~\cite{Studerus:2009ye,vonManteuffel:2012np}, and normalise
the scalar integrals as follows,
\begin{equation}\label{eq:intdef}
    I_{a_{1}, ..., a_{9}} = \int \bigg(\prod_{l=1}^2 (-q^2)^{-\epsilon} e^{\gamma_E \epsilon} \frac{\mathrm{d}^D k_l}{i\pi^{d/2}}\bigg) \prod_{i=1}^{9} D_i^{-a_i}\,,
\end{equation}
where $k_l$ indicates the loop momenta, $D_i \in \{P_i\}\cup\{N_i\}$ the propagators, and $\gamma_E = 0.577\dots$ the Euler-Mascheroni constant. 

All integrals can be reduced to a subset of master integrals using integration-by-parts
identities~\cite{Tkachov:1981wb,Chetyrkin:1981qh}. For the actual reduction, 
we use an implementation of the Laporta
algorithm~\cite{Laporta:2001dd} in the automated code \code{Kira2}~\cite{Maierhoefer:2017hyi,Klappert:2020nbg}, and express all integrals
directly in terms of the canonical basis~\cite{Henn:2013pwa} defined in~\cite{Gehrmann:2023etk}.
All integrals can be evaluated to arbitrary orders in the dimensional regularisation
parameter $\epsilon$ in terms for multiple polylogarithms (MPLs)~\cite{Goncharov:1998kja,Remiddi:1999ew,Gehrmann:2000zt,Vollinga:2004sn}
with alphabet
$\{y,z,1-y,1-z,y+z,1-y-z\}$.
We recall here that MPLs are defined as iterated integrals over logarithmic 
differential forms, whose arguments are linear rational functions,
\begin{align}
G(l_1,...,l_n;x) = \int_0^x \frac{dt}{t-l_1}G(l_2,...,l_n;t)\,,
\quad G(\underbrace{0,...,0}_n;x) = \frac{1}{n!} \log^n(x)
\end{align}
with $G(x) = 1$.
In this way, we can obtain an analytic expression for the bare form factors
in terms of MPLs in principle to arbitrary orders in $\epsilon$, though we limit ourselves to $\mathcal{O}(\epsilon^4)$ at one loop and $\mathcal{O}(\epsilon^2)$ at two loops.  We stress here that, up to this loop order and at variance with the $Vgq\bar q$ channel, the axial vector corrections for the $Vggg$ channel are individually well-defined in the Larin scheme for each flavour of massless quarks circulating in the loops, i.e.\ without having to sum over isospin doublets. This means that
our calculation is consistent without the corresponding contributions from the top quark, which would only start at $1/m_t^2$ in the $m_t \to \infty$ approximation.

\section{UV renormalisation and IR subtraction}\label{sec:UVandIR}
\subsection{UV renormalisation}
The form factors evaluated in the previous section contain UV and IR divergences. The former are removed by UV renormalisation. Denoting all unrenormalised quantities with a bar, we replace the bare coupling $\bar{\alpha}$ with the renormalised strong coupling $\alpha_{s}\equiv \alpha_{s}(\mu^2)$, evaluated at the renormalisation scale $\mu^2$,
\begin{equation}
    \bar{\alpha} \mu_{0}^{2\epsilon}S_{\epsilon} = \alpha_s \mu^{2\epsilon} \bigg[1-\frac{\beta_0}{\epsilon}\bigg(\frac{\alpha_s}{2 \pi}\bigg)+\bigg(\frac{\beta_0^2}{\epsilon^2}-\frac{\beta_1}{2 \epsilon}\bigg)\bigg(\frac{\alpha_s}{2 \pi}\bigg)^2+\mathcal{O}(\alpha_s^3)\bigg]\,,
\end{equation}
where $S_{\epsilon} = (4 \pi)^{\epsilon}\mathrm{e}^{-\epsilon \gamma_{E}}$ and $\mu_0^2$ is the mass parameter in dimensional regularization introduced to maintain a dimensionless coupling in the bare QCD Lagrangian density. For the remainder of the paper, we will set $\mu^2 = \mu_0^2 = q^2$.  The explicit form of the first two $\beta$-function coefficients $\beta_0,\, \beta_1$ reads 
\begin{align}
    \beta_0 &= \frac{11 C_A}{6} - \frac{2 T_R N_f}{3}, \\
    \beta_1 &= \frac{17 C_A^2}{6} - \frac{5 C_A T_R N_f}{3} -  C_F T_R N_f\,,
\end{align}
with the QCD colour factors
\begin{equation}
    C_A = N,\quad C_F = \frac{N^2-1}{2N},\quad T_R = \frac{1}{2}\,.
\end{equation}

Additionally for the $Vgq\bar{q}$ process, the renormalisation of the non-singlet axial vector
current requires multiplying the non-singlet contributions to the axial vector form factors $G_i$
by the non-singlet renormalisation constant, whose explicit value depends on the
the scheme used for treating $\gamma_5$. In the Larin scheme, it is known up to four loops~\cite{Chen:2021gxv}. We only require its value up to two loops in our calculation: 
\begin{align}
    Z_a^{n} = 1 -2 C_F \left( \frac{\alpha_s}{2 \pi} \right)
    + \left[ \frac{11}{2} C_F^2 - \frac{107}{36} C_F C_A
    + \frac{1}{18} C_F N_f \right]\left( \frac{\alpha_s}{2 \pi} \right)^2
    + \mathcal{O}(\alpha_s^3)\,.
\end{align}

 In contrast, the pure-singlet renormalisation constant starts only at order $\alpha_s^2$. Hence it enters in the $Vgq\bar{q}$ process already at two loops, while for $Vggg$ there is no tree amplitude and it would only contribute to the renormalisation of the three-loop amplitudes not computed here. For a single quark flavour, the pure-singlet renormalisation reads~\cite{Larin:1993tq}
 \begin{align}
    Z_a^{s} = C_FT_R \left( \frac{\alpha_s}{2 \pi} \right)^2
      \left[ \frac{3}{2\epsilon} + \frac{3}{4} \right] 
    + \mathcal{O}(\alpha_s^3)\,, \label{eq:zax}
\end{align}
where the pole stems from the UV renormalisation of the axial anomaly and the finite part is required to restore the axial Ward identities in dimensional regularization in the Larin scheme. 

In the following, we summarize the renormalisation formulas for both processes, with vector and axial vector couplings. For the gluonic channel, the vector boson can only attach to an internal quark line yielding just a singlet contribution. Due to the absence of a tree-level amplitude the renormalisation of the vector part is simply
\begin{align}
    F^{(1)}_{i} &= S_{\epsilon}^{-1}\bar{F}^{(1)}_{i},\\
    F^{(2)}_{i} &= S_{\epsilon}^{-2}\bar{F}^{(2)}_{i} - \frac{3\beta_0}{2\epsilon}S_{\epsilon}^{-1}\bar{F}^{(1)}_{i}\,,
    \end{align}
    while for the axial vector part an additional term 
    from the renormalisation of the axial vector 
    current is generated:
    \begin{align}
    G^{(1)}_{i} &= S_{\epsilon}^{-1}\bar{G}^{(1)}_{i},\\
    G^{(2)}_{i} &= S_{\epsilon}^{-2}\bar{G}^{(2)}_{i} - \frac{3\beta_0}{2\epsilon}S_{\epsilon}^{-1}\bar{G}^{(1)}_{i}-2 C_F \bar{G}_i^{(1)}\,.
    \label{eqn: axialrenorm1}
\end{align}
For the $Vgq\bar{q}$ channel with coupling to a vector current, we have first for the non-singlet contribution
\begin{align}
F_i^{(0),n} & = \bar{F}_i^{(0),n}\,, \nonumber \\
F_i^{(1),n}   &= S_\epsilon^{-1} \bar{F}_i^{(1),n} 
    - \frac{\beta_0}{2 \epsilon}  \bar{F}_i^{(0),n}\,, \nonumber \\
   F_i^{(2),n}   &= S_\epsilon^{-2} \bar{F}_i^{(2),n}  
    - \frac{3\beta_0}{2 \epsilon}  \bar{F}_i^{(1),n}  S_\epsilon^{-1} 
    - \left( \frac{\beta_1}{4 \epsilon} - \frac{3\beta_0^2}{8 \epsilon^2}\right) F_i^{(0),n}\,,  
    \end{align}
and second for the pure-singlet contribution
\begin{align}
  F_i^{(2),p} & = \bar{F}_i^{(2),p}  \,.
\end{align}
Still in the $Vgq\bar{q}$ channel but with the axial vector coupling, for the non-singlet part:
\begin{align}
G_i^{(0),n}  = &\; \bar{G}_i^{(0),n}   \,, \nonumber \\
   G_i^{(1),n}   =&\; S_\epsilon^{-1}  \bar{G}_i^{(1),n} 
    - \frac{\beta_0}{2 \epsilon}  \bar{G}_i^{(0),n} 
     -  2 C_F \bar{G}_i^{(0),n}\,, \nonumber \\
   G_i^{(2),n}   =&\; S_\epsilon^{-2} \bar{G}_i^{(2),n} 
    - \frac{3\beta_0}{2 \epsilon}  \bar{G}_i^{(1),n}   S_\epsilon^{-1} 
    - \left( \frac{\beta_1}{4 \epsilon} - \frac{3\beta_0^2}{8 \epsilon^2}\right)  \bar{G}_i^{(0),n}    \nonumber \\
     &-2 C_F   \bar{G}_i^{(1),n}   S_\epsilon^{-1}
     + \left( \frac{2 \beta_0}{\epsilon} + \frac{11}{2}C_F^2 
     - \frac{107}{36} C_F C_A
             + \frac{1}{18} C_F N_f\right)  \bar{G}_i^{(0),n} \,,
\end{align}
where the extra terms stem from the
 renormalisation of the axial vector current in the Larin scheme as described in~\cite{Gehrmann:2022vuk}. 
 Finally, the renormalisation of the pure-singlet axial vector form factors reads 
  \begin{align}
 G_i^{(1),p}  =&\;  S_\epsilon^{-1} \bar{G}_i^{(1),p}   \,, \nonumber \\
G_i^{(2),p} = &\; S_\epsilon^{-2}  \bar{G}_i^{(2),p} 
    - \left( \frac{ 3\beta_0}{2 \epsilon} \right) S_\epsilon^{-1} \bar{G}_i^{(1),p} 
     + C_FT_R  \left( \frac{3}{2\epsilon} + \frac{3}{4} \right)  \bar{G}_i^{(0),n} 
     -2 C_F \bar{G}_i^{(1),p}\, , 
        \label{eqn: axialrenorm2}
 \end{align}
for $N_f$ massless quark flavours running in the loops.

\subsection{IR subtraction}
The remaining IR singularities are universal in QCD and they depend only on the partons involved in the scattering process~\cite{Becher:2009qa, Catani:1998bh, Dixon:2009ur}. The method of subtracting these divergences is therefore identical to the one considered in~\cite{Gehrmann:2023etk} for the $Hggg$ and $Hgq\bar{q}$ processes.

Generally, the IR divergences in the renormalised two-loop form factors can be expressed in terms of the renormalised tree and one-loop form factors $\Phi^{(l)} = F_{i}^{(l)}, G_{i}^{(l)}$ multiplied by appropriate operators
\begin{align}
\Phi^{(0)}_{\text{finite}} & = \Phi^{(0)},\\
\Phi^{(1)}_{\text{finite}} & = \Phi^{(1)} - I_{\Phi}^{(1)} \, \Phi^{(0)} ,\label{eqn: perturbative IR subtraction one loop}\\
\Phi^{(2)}_{\text{finite}} & = \Phi^{(2)} - I_{\Phi}^{(2)} \, \Phi^{(0)}  - I_{\Phi}^{(1)} \, \Phi^{(1)}\,.
\label{eqn: perturbative IR subtraction two loops}
\end{align}
Specifically in the formalism of Soft-Collinear Effective Theory (SCET), the subtraction operators in \eqref{eqn: perturbative IR subtraction one loop}--\eqref{eqn: perturbative IR subtraction two loops} can be expressed as
\begin{align}
 I^{(1)} & = \mathcal{Z}^{(1)}\,,\\
 I^{(2)} & = \mathcal{Z}^{(2)} - \big(\mathcal{Z}^{(1)}\big)^2\,,
 \label{eqn: IR scet}
\end{align}
where the process dependence is implicit and the operators can be related to the relevant anomalous dimension coefficients via
\begin{align}
 \mathcal{Z}^{(1)} &= \frac{\Gamma'_0}{4 \epsilon^2} + \frac{\Gamma_0}{2 \epsilon}\,,\\
\mathcal{Z}^{(2)} &= \frac{{\Gamma_0'}^2}{32 \epsilon^4} +  \frac{\Gamma'_0}{8 \epsilon^3}\Big(\Gamma_0- \frac{3}{2}\beta_0 \Big) + \frac{\Gamma_0}{8 \epsilon^2}\Big( \Gamma_0 - 2 \beta_0\Big) + \frac{\Gamma'_1}{16 \epsilon^2} + \frac{\Gamma_1}{4 \epsilon}\,.
\label{eqn: Z scet}
\end{align}
Manifestly, the IR operators in SCET scheme have only pole terms but they affect the finite part of the amplitude when multiplied with the lower perturbative orders. For our two processes
\begin{align}
\Gamma^{ggg}_{n} &=- \frac{C_A}{2} \Big( L_{12} + L_{23} + L_{13} \Big) \gamma^{\text{cusp}}_n  + 3 \gamma_n^{g}\,, \\
\Gamma^{q\bar{q}g}_{n} &=  -C_F \, L_{12} \,  \gamma^{\text{cusp}}_n - \frac{C_A}{2} \Big(-L_{12} + L_{23} + L_{13} \Big) \gamma^{\text{cusp}}_n  + 2 \gamma_n^{q} + \gamma_n^{g}\,,
\label{eqn: GammaExpansion}
\end{align}
with 
\begin{align}
L_{ij} = \log{\Big(-\frac{\mu^2}{s_{ij}+ i \eta} \Big)}\,.
\end{align}

The cusp anomalous dimension $\gamma^{\rm cusp}$, quark anomalous dimension $\gamma_q$ and gluon anomalous dimension $\gamma_g$ have perturbative expansions
\begin{align}
\gamma^{\text{cusp}} = \sum_{n = 0}^{\infty} \gamma_n^{\text{cusp}} \Big( \frac{\alpha_s}{2 \pi} \Big)^{n+1}\,, \, \quad \, \gamma^{i} = \sum_{n = 0}^{\infty} \gamma_n^{i} \Big( \frac{\alpha_s}{2 \pi} \Big)^{n+1}
\end{align}
with $i=q,\bar{q},g$. We summarize here the coefficients up to $\mathcal{O}(\alpha_s^2)$ computed in~\cite{korchemksy:1987, vanneerven:1986, matsuura:1988sm, Harlander:2000}:
\begin{align}
    \gamma_0^{\text{cusp}} &= 2\,, \nonumber \\
    \gamma_1^{\text{cusp}} &= \left( \frac{67}{9} - \frac{\pi^2}{3} \right) C_A - \frac{10}{9}N_f\,, \nonumber \\
    \gamma_0^{q} &= -\frac{3}{2} \, C_F\,, \nonumber \\
    \gamma_1^{q} &= C_F^2 \left(  -\frac{3}{8} + \frac{\pi^2}{2}  - 6 \zeta_3\right) +
    C_F C_A \left(- \frac{961}{216} - \frac{11}{24} \pi^2  + \frac{13}{2}  \zeta_3  \right) + 
    C_F N_f \left(\frac{65}{108} + \frac{\pi^2}{12} \right)\,, \nonumber \\
\gamma_0^{g} &= - \beta_0\,, \nonumber \\
\gamma_1^{g} &= C_A^2 \left( -\frac{173}{27} + \frac{11}{72}\pi^2 + \frac{\zeta_3}{2} \right) + C_A N_f \left(\frac{32}{27} - \frac{\pi^2}{36}\right) + \frac{1}{2} C_F N_f\,.
\end{align}

\section{Helicity amplitudes}\label{sec:helicity}
In this section, we use the form factors from the previous chapter to construct the helicity amplitudes for the decay of an external off-shell vector and axial vector current into three partons. The case of $Vgq\bar{q}$ was described in detail in~\cite{Gehrmann:2022vuk} and here we focus on the gluonic channel. Subsequently we dress the helicity amplitudes with appropriate electroweak couplings to evaluate the decay of $\gamma^{*}, Z$ into three gluons, or $\gamma^{*}, Z, W^{\pm}$ into $gq\bar{q}$.

\subsection{Helicity amplitudes for an external vector and axial vector current}
Let us consider the production of the external current through lepton-antilepton annihilation
\begin{align}
     l^{+}(p_5)+l^{-}(p_6) \to V(p_4)\,
\end{align}
with fixed helicities of the initial state leptons. The associated left- and right-handed currents are
\begin{align}\label{eq:leptonspinhel1}
    C_L^\mu(p_5^{+},p_{6}^{-}) = [6 \gamma^\mu 5\rangle\,, \qquad C_R^\mu(p_5^{+},p_{6}^{-}) = [5 \gamma^\mu 6\rangle\,. 
\end{align}
There are only two independent helicity configurations of the three gluons, $(+++)$ and $(+--)$. In spinor-helicity formalism, for outgoing states and with a cyclic gauge choice, the $i^{\rm th}$ gluon's polarization vector is represented by
\begin{equation}\label{eq:gluonpol}
\epsilon_{i,-}^\mu = \frac{\langle i |\gamma^\mu |i+1 ]}{\sqrt{2} [i |i+1]}\,, \quad 
\epsilon_{i,+}^\mu = \frac{\langle i+1 |\gamma^\mu |i ]}{\sqrt{2} \langle i+1 |i \rangle}\,.
\end{equation}
Unlike in the computation of the form factors, here we are contracting with four-dimensional external states, and thus we need a four-dimensional representation of the axial vector,
\begin{equation}\label{eq:spinhelvA}
v_A^\mu \equiv \epsilon^{p_1 p_2 p_3 \mu} = \frac{1}{4} \Big[ [ 1 2 3 \gamma^\mu 1\rangle -\langle 1 2 3 \gamma^\mu 1] \Big]\,.
\end{equation}

In the two independent configurations $(+++L)$ and $(+--L)$, where the helicities of the external gluons are fixed, the dependence on the gluons' helicities can be summarized in a prefactor
\begin{align}
A_{\mu_1 \mu_2\mu_3\mu}\epsilon_{1,\lambda_1}^{\mu_1}\epsilon_{2,\lambda_2}^{\mu_2}\epsilon_{3,\lambda_3}^{\mu_3}\epsilon_{4}^{\mu_4} \quad\to\quad Sp(\lambda_1, \lambda_2, \lambda_3) \tilde{A}_{\mu}(p_1, p_2, p_3) \epsilon_{4}^{\mu}\,,
\end{align}
which in the two configurations reads
\begin{align}
Sp(+,+,+)&=\frac{1}{ \langle 1 2 \rangle \langle 2 3 \rangle \langle 1 3 \rangle }\,, \\
Sp(+,-,-)&=\frac{\langle 2 3 \rangle^3}{ \langle 1 2 \rangle \langle 1 3 \rangle }\,.
\end{align}
The number of independent spinor structures into which the stripped amplitude $\tilde{A}_{\mu}$ can be decomposed is therefore governed by the degrees of freedom of $\epsilon^{\mu}_4$. In fact, we can write this polarisation vector in terms of the momenta of just two of the three gluons, completed to a basis with two independent light-like vectors orthogonal to the gluon momenta,
\begin{align}
\epsilon_4^{\mu} = \alpha p_1^{\mu} + \beta p_2^{\mu} + \gamma \frac{\langle 2 \gamma^\mu 1 ]}{\langle 2 3 1 ]} + \delta \frac{\langle 1 \gamma^\mu 2 ]}{\langle 1 3 2 ]}\,.
\label{eq:eps4Decomposition}
\end{align}
In order to determine the coefficients $\alpha, \beta, \gamma, \delta$, we  evaluate the following contractions
\begin{alignat}{3}
&p_1 \cdot \epsilon_4 = \beta \frac{s_{12}}{2}\,, \qquad
&&p_2  \cdot \epsilon_4 = \alpha \frac{s_{12}}{2}\,, \nonumber\\
&\langle 2 \epsilon_4  1 ] = - 2 \delta \frac{s_{12}}{\langle 1 3 2 ]}\,, \qquad
&&\langle 1 \epsilon_4  2 ] = - 2 \gamma \frac{s_{12}}{\langle 2 3 1 ]}\,.
\label{eq:coefficientsEpsilon4}
\end{alignat}
The Lorenz gauge $\epsilon_4 \cdot p_4 = 0$ gives an additional constraint,
\begin{align}
\gamma = - \delta - \beta \left(\frac{s_{12}+s_{23}}{2}\right) - \alpha \left(\frac{s_{12}+s_{13}}{2}\right)\,.
\label{eq:LorenzGaugeConstraint}
\end{align}
We can now solve for the coefficients in \eqref{eq:coefficientsEpsilon4} and insert them back into \eqref{eq:eps4Decomposition} to get
\begin{align}\nonumber
\epsilon_4^\mu = &\frac{1}{s_{12}} ( p_{1\nu}  \epsilon_4^\nu) \Bigg( 2 p_2^\mu - \left(s_{12} + s_{23}\right)\frac{\langle 2 \mu 1 ]}{\langle 2 3 1]} \Bigg) +\\ \nonumber
 &\frac{1}{s_{12}} ( p_{2\nu}  \epsilon_4^\nu) \Bigg( 2 p_1^\mu - \left(s_{12} + s_{13}\right)\frac{\langle 2 \mu 1 ]}{\langle 2 3 1]} \Bigg) +\\ \label{eq:eps4 decomposition}
 & \frac{1}{2 s_{12}}( \langle 2 \nu 1 ]  \epsilon_4^\nu) \Bigg( \frac{\langle 1 3 2 ]}{\langle 2 3 1]} \langle 2 \mu 1] - \langle 1 \mu 2]\Bigg)\,.
\end{align}
Thanks to the constraint from the Lorenz gauge~\eqref{eq:LorenzGaugeConstraint}, the term proportional to $\langle1\nu2]\epsilon_4^\nu$ in the decomposition~\eqref{eq:eps4Decomposition} drops out.
Note that only the two gluon momenta $p_1, p_2$ carry open Lorentz indices on the right-hand side, whereas the polarization vectors $\epsilon_4$ are contracted.
The $\mu$-dependent bracketed terms will combine with $\tilde{A}^\mu(p_1, p_2, p_3)$, while the $\epsilon_4$ vectors will be substituted with the leptonic current in the chosen chirality configuration. From the contraction with the leptonic current, we observe three independent spinor structures, row by row in~\eqref{eq:eps4 decomposition}: $\langle 5 1 6 ]$, $\langle 5 2 6 ]$, and $ \langle25\rangle[16]\langle132]$. 

In summary, the contraction of the leptonic and hadronic currents yields a helicity amplitude of the form 
\begin{align}
{\rm M}_{\lambda_1 \lambda_2 \lambda_3 \lambda_{l_5}}(p_5,p_6;g_1,g_2,g_3) = \epsilon_{1,\lambda_1}^{\mu_1}\epsilon_{2,\lambda_2}^{\mu_2}\epsilon_{3,\lambda_3}^{\mu_3}   \, 
	A_{\mu_1 \mu_2\mu_3\mu}(g_1,g_2,g_3) C_{\lambda_{l_5}}^\mu(p_5,p_6)\,,
	 \label{eq:HAnocouplings}
\end{align}
which can be decomposed in terms of the three independent spinor structures by substituting the definitions \eqref{eq:leptonspinhel1}, \eqref{eq:gluonpol}, \eqref{eq:spinhelvA} into the general formula for the helicity amplitude \eqref{eq:HAnocouplings}. Through the tensor decomposition of the amplitude $A_{\mu_1 \mu_2\mu_3\mu}$ given in \eqref{eq:tensordecomposition}, the coefficients of the spinor structures in the helicity amplitude decomposition can be expressed entirely in terms of the form factors calculated in the previous section.

For the first independent helicity configuration we get
\begin{align}
{\rm M}_{+++L}^v &= \frac{1}{\sqrt{2}\langle12\rangle\langle23\rangle\langle13\rangle}\left(\alpha_1\langle516] + \alpha_2 \langle 5 2 6 ]+ \alpha_3 \langle13\rangle[23]\langle25\rangle[16] \right)\,,\\ 
{\rm M}_{+++L}^a &= \frac{1}{2\sqrt{2}\langle12\rangle\langle23\rangle\langle13\rangle}\left(\beta_1\langle516] + \beta_2 \langle 5 2 6 ]+ \beta_3 \langle13\rangle[23]\langle25\rangle[16] \right)\,, 
\end{align}
where the helicity coefficients $\alpha_i$ and $\beta_i$ 
are related to the form factors as follows:
\begin{align}
  \alpha_1 &= s_{12} s_{13}\left(-\frac{s_{23}}{2}F_{1}+F_3-F_8\right)-s_{23} \left(s_{13}F_{5}+s_{12}F_{10}\right)+s_{23}\left(s_{23}+s_{12} \right)F_4  \nonumber
   \\ \nonumber &\phantom{ =\,\,}
   +2 F_{12} \left(2 s_{12}+s_{23}\right)
 \\  \nonumber
 \alpha_2 &=  s_{12} s_{23}\left(-\frac{s_{13}}{2}F_{2}+F_4-F_{11}\right)-s_{13} \left(s_{23}F_{6}+s_{12}F_{9}\right)+s_{13}\left(s_{13}+s_{12}\right)F_3 
  \\ \nonumber &\phantom{ =\,\,}
  +2 F_{12} \left(2 s_{12}+s_{13}\right)\\
  \alpha_3 &=  - s_{13}F_3-s_{23}F_4 - s_{12}F_7-6 F_{12} \label{eq:alpha3}
\end{align}
and
\begin{align}
   \beta_1&=  -\frac{s_{12}s_{23}s_{13}}{2}\left(-(s_{12}+s_{23})G_1+
   s_{13}G_{2}+ s_{23}G_{4}+s_{12}G_{7}+6G_{11}\right) \nonumber \\\nonumber &\phantom{ =\,\,}
   +s_{13} s_{23}\left(2 s_{23}+3 s_{12} \right)G_6 +s_{13}s_{12}\left(2  s_{12}+ s_{23}\right)G_9 +2s_{23}s_{12} \left(s_{23} + s_{12}\right) G_{10}\\ \nonumber
   \beta_2&=-\frac{s_{12}s_{23}s_{13}}{2}\left(-(s_{12}+s_{13})G_1+s_{13}G_{3}+ s_{23}G_{5}+s_{12}G_{8}+6G_{12}\right)  \\\nonumber &\phantom{ =\,\,}
   +s_{23} s_{13}\left(2 s_{13}+3 s_{12} \right)G_6  +2 s_{12}s_{13}\left(s_{12}+s_{13} \right)G_9 +s_{23} s_{12}\left(2 s_{12}+s_{13}\right)G_{10} \\
   \beta_3 &= -s_{12} s_{13}\left(s_{23}G_{1}+4G_{9}\right)-4  s_{23} s_{13}G_6-4 s_{12} s_{23}G_{10} \,. \label{eq:beta3}
\end{align}
Next, for the $(+--L)$ configuration, the helicity amplitudes are
\begin{align}
{\rm M}_{+--L}^v &= \frac{\langle23\rangle^3}{\sqrt{2}\langle12\rangle\langle13\rangle}\left(\gamma_1\langle516] + \gamma_2 \langle 5 2 6 ]+ \gamma_3 \langle13\rangle[23]\langle25\rangle[16]  \right)\,,\\ 
{\rm M}_{+--L}^a &= \frac{\langle23\rangle^3}{2\sqrt{2}\langle12\rangle\langle13\rangle}\left(\delta_1\langle516] + \delta_2 \langle 5 2 6 ]+ \delta_3 \langle13\rangle[23]\langle25\rangle[16] \right)\,,
\end{align}
with
\begin{align}
    \gamma_1 &= -\frac{s_{13}}{s_{23}}\left(\frac{s_{12}}{2}F_{1}+F_3+F_5\right) \nonumber
   \\
   \gamma_2 &=  -\frac{s_{13}}{s_{23}}\left(\frac{s_{12}}{2}F_{2}+F_4+F_6\right) \nonumber
   \\
   \gamma_3 &= \frac{1}{s_{23}^2}\left(s_{13}F_{3}+s_{23}F_{4}-s_{12}F_{7}-2F_{12}\right) \label{eq:gamma3}
\end{align}
and
\begin{align}
   \delta_1 &= \frac{s_{13}s_{12}}{2s_{23}}\left((s_{12}+s_{23})G_1+G_{2}s_{13} +G_{4}s_{23} - G_{7}s_{12}-2G_9 -2G_{11}\right)\nonumber\\\nonumber &\phantom{ =\,\,}
+\frac{G_6 s_{13} 
\left(s_{12}+2 s_{23}\right)}{s_{23}}\\
  \delta_2 &= \frac{s_{13}s_{12}}{2s_{23}}\left((s_{12}+s_{13})G_1+G_{3}s_{13} +G_{5}s_{23} - G_{8}s_{12}-2G_{10} -2G_{12}\right)\nonumber\\\nonumber &\phantom{ =\,\,}
  +\frac{G_6 s_{13}\left(2 s_{13}+s_{12}\right)}{s_{23}} \\  
\delta_3 &= \frac{1}{s_{23}^2}\left(-G_{1}s_{12}s_{13}s_{23}-4G_{6}s_{13}s_{23}+2G_{9}s_{12}s_{13}+2G_{10}s_{12}s_{23}\right)\,. \label{eq:delta3}
\end{align}
The relations \eqref{eq:alpha3}, \eqref{eq:beta3}, \eqref{eq:gamma3}, and \eqref{eq:delta3} between form factors and helicity amplitude coefficients are also provided in computer-readable format in the ancillary files.

In the $Vggg$ process, there are a total of $2^3\times 2 =16$ different helicity configurations for each of the vector and axial vector contributions. From the above expressions for 
${\rm M}_{+--L}(p_5,p_6;g_1,g_2,g_3)$ and ${\rm M}_{+++L}(p_5,p_6;g_1,g_2,g_3)$, all the other helicity amplitudes can be obtained
by CP conjugation and permutations of the external legs:
\begin{align}
 {\rm M}^{a,v}_{-+-L}(p_5,p_6;g_1,g_2,g_3) &= {\rm M}^{a,v}_{+--L}(p_5,p_6;g_2,g_1,g_3)\,,   \nonumber\\
 {\rm M}^{a,v}_{--+L}(p_5,p_6;g_1,g_2,g_3) &= {\rm M}^{a,v}_{+--L}(p_5,p_6;g_3,g_2,g_1)  \,, \nonumber\\
 {\rm M}^{v}_{++-L}(p_5,p_6;g_1,g_2,g_3) &= [{\rm M}^{v}_{+--L}(p_6,p_5;g_3,g_2,g_1)]^*\,,  \nonumber\\
 {\rm M}^{v}_{+-+L}(p_5,p_6;g_1,g_2,g_3) &= [{\rm M}^{v}_{+--L}(p_6,p_5;g_2,g_1,g_3)]^* \,,  \nonumber\\
 {\rm M}^{v}_{-++L}(p_5,p_6;g_1,g_2,g_3) &= [{\rm M}^{v}_{+--L}(p_6,p_5;g_1,g_2,g_3)]^* \,,  \nonumber\\
 {\rm M}^{v}_{---L}(p_5,p_6;g_1,g_2,g_3) &= [{\rm M}^{v}_{+++L}(p_6,p_5;g_1,g_2,g_3)]^*\,,  \nonumber\\
{\rm M}^{a}_{++-L}(p_5,p_6;g_1,g_2,g_3) &= -[{\rm M}^{a}_{+--L}(p_6,p_5;g_3,g_2,g_1)]^*\,,  \nonumber\\
 {\rm M}^{a}_{+-+L}(p_5,p_6;g_1,g_2,g_3) &= -[{\rm M}^{a}_{+--L}(p_6,p_5;g_2,g_1,g_3)]^* \,,  \nonumber\\
 {\rm M}^{a}_{-++L}(p_5,p_6;g_1,g_2,g_3) &= -[{\rm M}^{a}_{+--L}(p_6,p_5;g_1,g_2,g_3)]^* \,,  \nonumber\\
 {\rm M}^{a}_{---L}(p_5,p_6;g_1,g_2,g_3) &= -[{\rm M}^{a}_{+++L}(p_6,p_5;g_1,g_2,g_3)]^* .  \label{eq:HelAmplggg}
 \end{align}
The corresponding amplitudes for the right-handed leptonic current are obtained by interchanging $p_5 \leftrightarrow p_6$.
Note that the complex conjugation operation acts only on the spinor structures and not on the coefficients $\alpha_i,\ldots,\delta_i$.

The equivalent expressions for the independent helicity amplitudes ${\rm M}_{L+L}$ and ${\rm M}_{L-L}$ in terms of the form factors for the process $V \to q\bar{q}g$ were given in section 5.1 of~\cite{Gehrmann:2022vuk}. In the same notation (where the first subscript belongs to the quark and the last to the lepton), the remaining two helicity amplitudes can be obtained via
\begin{align}
 {\rm M}^{v}_{R+L}(p_5,p_6;\bar{q}_1,q_2,g_3) &= [{\rm M}^{v}_{L-L}(p_6,p_5;\bar{q}_1,q_2,g_3)]^{*}\,,   \nonumber\\
 {\rm M}^{v}_{R-L}(p_5,p_6;\bar{q}_1,q_2,g_3) &= [{\rm M}^{v}_{L+L}(p_6,p_5;\bar{q}_1,q_2,g_3) ]^{*}\,,   \nonumber\\
  {\rm M}^{a}_{R+L}(p_5,p_6;\bar{q}_1,q_2,g_3) &= -[{\rm M}^{a}_{L-L}(p_6,p_5;\bar{q}_1,q_2,g_3)]^{*}\,,   \nonumber\\ \label{eq:HelAmplgqq}
 {\rm M}^{a}_{R-L}(p_5,p_6;\bar{q}_1,q_2,g_3) &= -[{\rm M}^{a}_{L+L}(p_6,p_5;\bar{q}_1,q_2,g_3) ]^{*}
\end{align}
and are related to the amplitudes with a right-handed leptonic current in the same way.

The finite remainders for vector and axial vector helicity coefficients
$\Omega_i = \left\{ \alpha_i,\gamma_i \right\}$ and 
$\Lambda_i = \left\{\beta_i, \delta \right\}$ can be expanded as series in the strong coupling
\begin{align}
\Omega_i &= \Omega_i^{(0)} + \left( \frac{\alpha_s}{2 \pi}\right) \Omega_i^{(1)} 
     + \left( \frac{\alpha_s}{2 \pi}\right)^2 \Omega_i^{(2)} +\mathcal{O}(\alpha_s^3)\,, \nonumber \\
\Lambda_i &= \Lambda_i^{(0)} + \left( \frac{\alpha_s}{2 \pi}\right) \Lambda_i^{(1)} 
     + \left( \frac{\alpha_s}{2 \pi}\right)^2 \Lambda_i^{(2)} +\mathcal{O}(\alpha_s^3)\,.  \label{eq:OmLam}   
\end{align}
We report $\Omega_i$ and $\Lambda_i$ through to two loops in the ancillary files.

\subsection{Helicity amplitudes for Standard Model vector bosons}
We are now ready to build the helicity amplitudes for a
Standard Model vector boson $V$ connecting either to a $q\bar{q}g$ or a $ggg$ system and to a lepton-antilepton pair.
Let us write the coupling of the vector boson $V$ to two fermions $f_1 f_2$ 
in two equivalent ways,
\begin{align}
    -i e\, \Gamma_\mu^{V f_1 f_2} =& 
    - i \sqrt{4 \pi \alpha}\, \left[ v^V_{f_1 f_2} \gamma^\mu + a^V_{f_1 f_2} \gamma^\mu \gamma_5 \right] \nonumber \\
    =&-i \sqrt{4 \pi \alpha}\, \left[L_{f_1 f_2}^V \gamma^\mu \left( \frac{1-\gamma_5}{2} \right) + R_{f_1 f_2}^V \gamma^\mu \left( \frac{1+\gamma_5}{2} \right)\right]\,,
    \label{eq:ewvertex}
\end{align}
where
\begin{align}
L_{f_1 f_2}^V = v^V_{f_1 f_2} - a^V_{f_1 f_2} \,, \quad
R_{f_1 f_2}^V = v^V_{f_1 f_2} + a^V_{f_1 f_2}\,. 
\end{align}
For the three types of vector bosons ($W^{\pm}$ only in the quark-antiquark channel), we have
\begin{align}
    L_{f_1 f_2}^\gamma = R_{f_1 f_2}^\gamma = - e_{f_1} \delta_{f_1 f_2}
    \end{align}
\begin{align}
    L_{f_1 f_2}^Z = \frac{I_3^{f_1} - \sin^2{\theta_w} e_{f_1}}{\sin{\theta_w} \cos{\theta_w}} \delta_{f_1 f_2} \,, \qquad
    R_{f_1 f_2}^Z = -\frac{\sin{\theta_w} e_{f_1}}{\cos{\theta_w}} \delta_{f_1 f_2} \,,
\end{align}    
    \begin{align}
    L_{f_1 f_2}^W = \frac{\epsilon_{f_1,f_2}}{\sqrt{2}\sin{\theta_w}} \,, \qquad
    R_{f_1 f_2}^W = 0 \,.
\end{align}  
In the formulas above, $\alpha$ is the electroweak coupling constant,
$\theta_w$ is the Weinberg angle, $I_3 = \pm 1/2$ is the third component
of the weak isospin and the charges $e_i$
are measured in terms of the fundamental electric charge $e>0$.
Moreover, $\epsilon_{f_1,f_2}=1$ if $f_1 \neq f_2$ but belonging to the same
isospin doublet and zero otherwise. 
Finally, the expression for the propagator of the vector boson $V$ in Lorenz gauge, with momentum $q$ and mass $m_V$, is
\begin{equation}
    P_{\mu \nu}(q,m_V) = \frac{i \left( -g_{\mu \nu} + \frac{q_\mu q_\nu}{q^2}  \right)}{D\left(q^2,m_V^2\right)}
    \label{eq:propagatorV}
\end{equation}
with
\begin{align}
D\left(q^2,m_V^2\right) = q^2 - m_V^2 + i \Gamma_V m_V\,.
\end{align}

From (\ref{eq:ewvertex}), we can immediately read off 
the coupling factors of all non-singlet contributions: $v^V_{f_1 f_2}$ for the vector form factors and $a^V_{f_1 f_2}$ for the axial vector form factors. 
In the pure-singlet contributions to the $Vgq\bar{q}$ channel, and in the entire amplitude in the gluonic channel, the vertex (\ref{eq:ewvertex}) is coupled to an internal quark loop, thus requiring a summation over the internal quark flavours. In the vector case, this summation amounts to an overall factor
\begin{align}
    N_{f,\mathcal{\gamma}}^{v} =\sum_q e_q \,, \qquad
     N_{f,Z}^{v} =\sum_q \frac{\left( L_{qq}^Z  + R_{qq}^Z\right)}{2}\,, \qquad
     N_{f,W}^{v} = 0 \label{eq:Nfv}
\end{align}
with the sums running over the active quark flavours. The last identity in~\eqref{eq:Nfv} 
is a consequence of charge conservation. There can be a contribution from the 
axial vector coupling 
 only in the case
of the production of a $Z$ boson.
In the axial vector case, the summation must be performed 
over complete quark isospin doublets. For mass-degenerate quarks in the doublet, the up-type and down-type contributions cancel identically in the sum. In case of 
a mass-splitting in the doublet, the axial vector pure-singlet contribution is obtained as the difference between up-type and down-type quark contributions in the loop, multiplied by a coupling factor 
\begin{align}
N_{f,\gamma}^{a} = 0\,, \qquad 
     N_{f,Z}^{a} =  \frac{1}{4 \sin{\theta_w} \cos{\theta_w}}  \,,
     \qquad
     N_{f,W}^{a} = 0\,. \label{eq:Nfa}
\end{align}

With these definitions, we write the helicity amplitudes in the $Vgq\bar{q}$ channel as
\begin{align}
	\mathcal{M}_{L+L}^V &=- 
 \frac{i \sqrt{4 \pi \alpha_s} (4 \pi \alpha)\, L_{l_5 l_6}^{V} }{\sqrt{2} \,D(p_{56}^2,m_V^2)}\, \mathbb{T}^{a}_{ij}
	 \bigg[ \langle 1 2 \rangle [1 3]^2 \Big( A_1 \langle 536] 
	+ A_2 \langle 526] \Big) \nonumber\\
       &\phantom{\hspace{130pt}}+ A_3 \langle 25 \rangle [13] [36] \bigg] \,,
	 \label{eq:HALmL} \\
	\mathcal{M}_{L-L}^V &=  - 
  \frac{i \sqrt{4 \pi \alpha_s} (4 \pi \alpha)\, L_{l_5 l_6}^{V}}{\sqrt{2}\, D(p_{56}^2,m_V^2)}\,  \mathbb{T}^{a}_{ij}
  \bigg[ 
  \langle 23 \rangle^2 [1 2] \Big( B_1 \langle 536] 
	+ B_2 \langle 516] \Big) \nonumber\\
        &\phantom{\hspace{130pt}}+ B_3 \langle 23 \rangle \langle 35 \rangle [16]  
\bigg]\,, \label{eq:HALpL}
 \end{align}
with $p_{56} = p_5 + p_6$. The terms $A_i, B_i$ can be related to the $l$-loop helicity amplitude coefficients $\alpha_i,\ldots,\delta_i$ for the $gq\bar{q}$ channel given in~\cite{Gehrmann:2022vuk} via
\begin{align}
A_i^{(l)} &=  L_{q_1 q_2}^V \alpha_i^{(l),n} 
+ N_{f,V}^v \alpha_i^{(l),p} + N_{f,V}^a \left( \beta_i^{(l),p0} - \beta_i^{(l),pm} \right)\,, \\
B_i^{(l)} &=  L_{q_1 q_2}^V \gamma_i^{(l),n} 
+ N_{f,V}^v \gamma_i^{(l),p} + N_{f,V}^a \left( \delta_i^{(l),p0} - \delta_i^{(l),pm} \right) \,,
\end{align}
where we used the fact that the non-singlet vector and axial vector
parts are equal after UV renormalisation and IR subtraction. Following~\cite{Gehrmann:2022vuk}, $n$ is the non-singlet part while $p0$ and $pm$ represent the massless and massive pure-singlet contributions. The latter are provided in~\cite{Gehrmann:2022vuk} and the conversion between the SCET scheme employed here and the Catani IR subtraction scheme which was used in the reference, is explained in~\cite{Gehrmann:2023etk}.

In the gluonic channel we have 
\begin{align}
	  \mathcal{M}_{+++L}^{V} &=  - 
  \frac{i \sqrt{4 \pi \alpha_s} (4 \pi \alpha)\, L_{l_5 l_6}^{V}}{\, \sqrt{2}\,D(p_{56}^2,m_V^2)}\,  \Bigg[
  \frac{1}{\langle12\rangle\langle23\rangle\langle13\rangle}\Big(C_1\langle516] + C_2 \langle 5 2 6 ]\nonumber\\
  &\phantom{\hspace{130pt}}+ C_3 \langle13\rangle[23]\langle25\rangle[16] \Big)\Bigg]\,,\\
   \mathcal{M}_{+--L}^{V}  &=  - 
  \frac{i \sqrt{4 \pi \alpha_s} (4 \pi \alpha)\, L_{l_5 l_6}^{V}}{\sqrt{2}\, D(p_{56}^2,m_V^2)}\,  \Bigg[
  \frac{\langle23\rangle^3}{\langle12\rangle\langle13\rangle}\Big(D_1\langle516] + D_2 \langle 5 2 6 ]\nonumber\\
  &\phantom{1\hspace{130pt}}+ D_3 \langle13\rangle[23]\langle25\rangle[16]  \Big)\Bigg]\,. \label{eq:HALpL}
 \end{align}
Here the $l$-loop coefficients can be written as 
\begin{align}
C_i^{(l)} &=  
d^{a_1 a_2 a_3} N_{f,V}^v \alpha_i^{(l),p} + \frac{1}{2}f^{a_1 a_2 a_3} N_{f,V}^a \beta_i^{(l),p} \,, \\
D_i^{(l)} &=  
d^{a_1 a_2 a_3} N_{f,V}^v \gamma_i^{(l),p} + \frac{1}{2} f^{a_1 a_2 a_3} N_{f,V}^a \delta_i^{(l),p} \,.
\end{align}
We stress that the remaining helicity amplitudes can be obtained from the ones
given above through a CP transformation, as described in~\eqref{eq:HelAmplggg} and~\eqref{eq:HelAmplgqq}, by swapping angle and square brackets and changing all relevant couplings, but leaving the functional part of the coefficients unchanged. 

\subsection{Analytic continuation}\label{sec:continuations}
In view of phenomenology studies relevant to LHC physics, we perform the analytic continuation of the decay amplitudes to the other regions where the electroweak boson is produced in addition to a jet from the scattering of two partons.
Starting from the decay of a $V$ boson into three gluons, $V \rightarrow ggg$, and into a quark-antiquark pair and gluon, $V \rightarrow q\bar{q} g$, we are interested in the regions where a $V$ is produced together with a parton: $gg \rightarrow Vg$, $ q g \rightarrow V q$, $\bar{q} g \rightarrow V \bar{q}$ and $q \bar{q} \rightarrow V g $.

The general strategy for performing the analytic continuation of the MPLs appearing in $2\to 2$ scattering involving 4-point functions with one off-shell leg and massless propagators was outlined in  detail in~\cite{Gehrmann:2002zr} and summarised in~\cite{Gehrmann:2023etk}. With reference to figure~1 in~\cite{Gehrmann:2002zr}, the goal is to analytically continue MPLs from the decay region (1a) to the three production regions, (2a), (3a), (4a).
In~\cite{Gehrmann:2023etk}, it is suggested to first consider suitable crossings of the amplitude in the decay region and in a second step to continue these crossed amplitudes exclusively to the region (3a), which is eventually equivalent to the direct approach. We~follow the same strategy here.

Unlike in the previous sections, where the vector boson was produced from an incoming lepton current, in the production regions we need to consider the boson's decay into a lepton pair, which amounts to swapping $L$ and $R$ compared to~\eqref{eq:leptonspinhel1},
\begin{align}\label{eq:leptonspinhel}
    C_R^\mu(p_5^{+},p_{6}^{-})\Big|_{\rm out} = [6 \gamma^\mu 5\rangle\,, \qquad C_L^\mu(p_5^{+},p_{6}^{-})\Big|_{\rm out} = [5 \gamma^\mu 6\rangle\,. 
\end{align}
Since the amplitudes with a left- and right-handed current are related by a simple $p_5 \leftrightarrow p_6$ swap, we explicitly provide only one copy, as indicated in table~\ref{tab:crossings}. The amplitudes in this section therefore only depend on the helicities of the three partons and we adjust our notation accordingly. 
For the case of $Vggg$, the two independent amplitudes in the decay kinematics are
\begin{align}
\mathcal{M}^{+++} : \quad \quad  V &\rightarrow g_+(p_1) + g_+(p_2) + g_+(p_3)\,,  \\
\mathcal{M}^{+--} : \quad \quad V &\rightarrow g_+(p_1) + g_-(p_2) + g_-(p_3)\,.
\end{align}
In the production regions, there are eight different helicity configurations. Thanks to parity symmetry, we can limit ourselves to considering just four of them (see the left column of table~\ref{tab:crossings}) and relating them to the other four (right column).
\begin{table}[t]
\begin{center}
\renewcommand{\arraystretch}{1.2}
\begin{tabular}{c c}
   \multicolumn{1}{c}{Process} & \multicolumn{1}{c}{Parity-related} \\
   \hline
  $\mathcal{M}^{--,+}_{gg,g}: \, \, g_-(p_1) + g_-(p_3) \rightarrow V_R + g_+(p_2)\, \quad $    &   $\quad \quad \,g_+(p_1) + g_+(p_3) \rightarrow V_L + g_-(p_2)$ \\
   $\mathcal{M}^{-+,-}_{gg,g}: \, \, g_-(p_1) + g_+(p_3) \rightarrow V_R + g_-(p_2)\, \quad$    &   $\quad \quad \,g_+(p_1) + g_-(p_3) \rightarrow V_L + g_+(p_2)$ \\
  $\mathcal{M}^{++,+}_{gg,g}: \, \, g_+(p_1) + g_+(p_3) \rightarrow V_R + g_+(p_2)\, \quad$    &   $\quad \quad \,g_-(p_1) + g_-(p_3) \rightarrow V_L + g_-(p_2)$ \\
   $ \mathcal{M}^{+-,+}_{gg,g}: \, \, g_+(p_1) + g_-(p_3) \rightarrow V_R + g_-(p_2)\, \quad$    &   $\quad \quad \,g_-(p_1) + g_+(p_3) \rightarrow V_L + g_+(p_2)$ \\ 
\end{tabular}
\end{center}
\caption{The helicity configurations for V$+$jet production in the $ggg$ channel. We provide expressions for those in the left column. A comma is used to separate the helicities of initial- and final-state partons. The subscripts $V_{L,R}$ indicate whether the boson decays into an outgoing left- or right-handed leptonic curent.
\label{tab:crossings}}
\end{table}
As noted in~\cite{Gehrmann:2023etk}, in the continuation to the region (3a), the momenta $p_1$ and $p_3$ are always in the initial state. We computed the first three amplitudes in the left column with a combination of crossings, analytic continuation and time reversal as follows:
\begin{align}
\mathcal{M}^{+++}_{ggg}\, &\xrightarrow[]{\text{to (3a)}} \,  \mathcal{M}^{--,+}_{gg,g} \,,\nonumber\\
\mathcal{M}^{+--}_{ggg}\, &\xrightarrow[]{\text{to (3a)}} \,  \mathcal{M}^{-+,-}_{gg,g} \,,\nonumber\\
\mathcal{M}^{+--}_{ggg}\, \xrightarrow[]{p_1 \leftrightarrow p_2} \,& \mathcal{M}^{-+-}_{ggg}\, \xrightarrow[]{\text{to (3a)}} \,  \mathcal{M}^{++,+}_{gg,g}\,.
\end{align}
The last amplitude $\mathcal{M}^{+-,-}_{gg,g}$ can be easily obtained from $\mathcal{M}^{-+,-}_{gg,g}$ under the crossing $p_1 \leftrightarrow p_3$. All amplitudes in the production region are written in terms of the new variables $(u,v)$ defined as
\begin{align}
y = \frac{1}{v}\,, \quad \quad z = -\frac{u}{v}\,.
\end{align}
The suggested crossing $p_1 \leftrightarrow p_3$ then implies the mapping $v \rightarrow v$ and $ u \rightarrow 1 - u - v$ for reaching $\mathcal{M}^{+-,-}_{gg,g}$ from $\mathcal{M}^{-+,-}_{gg,g}$. Under this transformation, no branch cut is crossed and no new analytic continuation is needed, so
\begin{align}
\mathcal{M}^{+-,-}_{gg,g}(v,u) =  \mathcal{M}^{-+,-}_{gg,g}(v,u)\vert_{u \rightarrow 1-u-v}\,.
\end{align}

Let us now consider the process $V\to gq\bar{q}$. In the decay region, we computed the same two helicity configuration as in~\cite{Gehrmann:2022vuk}, namely
\begin{align}
\mathcal{M}^{RL+} : \quad \quad  V &\rightarrow \bar{q}_R(p_1) + q_L(p_2) + g_+(p_3)\,,  \\
\mathcal{M}^{RL-} : \quad \quad  V &\rightarrow \bar{q}_R(p_1) + q_L(p_2) + g_-(p_3)\,,
\end{align}
formerly denoted $\mathcal{M}^{V}_{L+L}$ and $\mathcal{M}^{V}_{L-L}$, respectively\footnote{In the old notation, the first subscript referred to the quark, the last to the lepton. Here we drop the lepton subscript and label in order of the external momenta: anti-quark, quark, gluon.}. In other words, we again consider only one copy of the amplitude with respect to the leptonic current, while introducing a label for the handedness of the antiquark which might no longer be fixed by that of the quark in the production region.
There are 12 non-zero independent helicity configurations. Again we give solutions for 6 of them, the others being related by parity, as indicated in table~\ref{tab:crossQ}.

\begin{table}[t]
\begin{center}
   \renewcommand{\arraystretch}{1.2}
\begin{tabular}{c c}
   \multicolumn{1}{c}{Process} & \multicolumn{1}{c}{Parity-related} \\
   \hline
   $\mathcal{M}^{L-,L}_{\bar{q}g,\bar{q}}: \, \, \bar{q}_L(p_1) + g_-(p_3) \rightarrow V_{R} + \bar{q}_L(p_2)\, \quad$    &   $\quad \quad \,q_R(p_1) + g_+(p_3) \rightarrow V_L + q_R(p_2)$ \\ 
      $\mathcal{M}^{L+,L}_{\bar{q}g,\bar{q}}: \, \, \bar{q}_L(p_1) + g_+(p_3) \rightarrow V_{R} + \bar{q}_L(p_2)\, \quad$    &   $\quad \quad \,q_R(p_1) + g_-(p_3) \rightarrow V_L + q_R(p_2)$ \\ 
  $\mathcal{M}^{R-,R}_{qg,q}: \, \, q_R(p_1) + g_-(p_3) \rightarrow V_R + q_R(p_2)\, \quad $    &   $\quad \quad \,\bar{q}_L(p_1) + g_+(p_3) \rightarrow V_L + \bar{q}_L(p_2)$ \\
  $\mathcal{M}^{R+,R}_{qg,q}: \, \, q_R(p_1) + g_+(p_3) \rightarrow V_R + q_R(p_2)\, \quad $    &   $\quad \quad \,\bar{q}_L(p_1) + g_-(p_3) \rightarrow V_L + \bar{q}_L(p_2)$ \\
  $\mathcal{M}^{LR,+}_{\bar{q}q,g}: \, \, \bar{q}_L(p_1) + q_R(p_3) \rightarrow V_R + g_+(p_2)\, \quad $    &   $\quad \quad \,q_R(p_1) + \bar{q}_L(p_3) \rightarrow V_L + g_-(p_2)$ \\ 
  $\mathcal{M}^{LR,-}_{\bar{q}q,g}: \, \, \bar{q}_L(p_1) + q_R(p_3) \rightarrow V_R + g_-(p_2)\, \quad $    &   $\quad \quad \,q_R(p_1) + \bar{q}_L(p_3) \rightarrow V_L + g_+(p_2)$ \\ 
\end{tabular}
\end{center}
\caption{The helicity configurations for V$+$jet production in the $gq\bar{q}$ channel. We provide expressions for those in the left column. A comma is used to separate the helicities of initial- and final-state partons. The subscripts $V_{L,R}$ indicate whether the boson decays into an outgoing left- or right-handed leptonic curent.\label{tab:crossQ}}
\end{table}
The six amplitudes in the left column were computed with a combination of crossings, analytic continuations and time invesion as follows:
\begin{align}
\mathcal{M}^{RL+}_{\bar{q}qg}\, &\xrightarrow[]{\text{to (3a)}} \,  \mathcal{M}^{L-,L}_{\bar{q}g,\bar{q}} \,,\nonumber\\ 
\mathcal{M}^{RL-}_{\bar{q}qg}\, &\xrightarrow[]{\text{to (3a)}} \,  \mathcal{M}^{L+,L}_{\bar{q}g,\bar{q}} \,,\nonumber\\ 
\mathcal{M}^{RL+}_{\bar{q}qg}\, \xrightarrow[]{p_1 \leftrightarrow p_2} \,&  \mathcal{M}^{LR+}_{q\bar{q}g}\, \xrightarrow[]{\text{to (3a)}} \,  \mathcal{M}^{R-,R}_{qg,q}  \,,\nonumber\\ 
\mathcal{M}^{RL-}_{\bar{q}qg}\, \xrightarrow[]{p_1 \leftrightarrow p_2} \,&  \mathcal{M}^{LR-}_{q\bar{q}g}\, \xrightarrow[]{\text{to (3a)}} \,  \mathcal{M}^{R+,R}_{qg,q} \,, \nonumber\\ 
\mathcal{M}^{RL+}_{\bar{q}qg}\, \xrightarrow[]{p_2 \leftrightarrow p_3} \,&  \mathcal{M}^{R+L}_{\bar{q}gq}\, \xrightarrow[]{\text{to (3a)}} \,  \mathcal{M}^{LR,+}_{\bar{q}q,g} \,,\nonumber\\
\mathcal{M}^{RL-}_{\bar{q}qg}\, \xrightarrow[]{p_2 \leftrightarrow p_3} \,&  \mathcal{M}^{R-L}_{\bar{q}gq}\, \xrightarrow[]{\text{to (3a)}} \,  \mathcal{M}^{LR,-}_{\bar{q}q,g} \,.
\end{align}

\section{Checks and results}\label{section:results}
We performed a number of validations on 
the form factors and the resulting helicity amplitudes.

The first check concerns the renormalisation of the axial vector coupling. The Larin prescription for treating $\gamma_5$ used in this work does not by itself correctly recover the axial vector current defined as
\begin{align}
J_5^\mu=\overline{\psi}\gamma^\mu\gamma_5\psi\,.
\end{align}
The correct current can be restored by introducing a finite renormalisation constant $Z_a$ on top of the standard UV renormalisation constants in the $\overline{ {\rm MS}}$-scheme. 
While the standard renormalisation constant is fixed by the cancellation of UV poles in the diagrammatic expansion for the form factors, $Z_a$ only affects the finite part and is needed to restore the correct axial Ward identities~\cite{Larin:1993tq}.

The finite renormalisation $Z_a$
is determined by requiring that the 
divergence of the 
correctly renormalised 
axial vector current satisfies the
axial anomaly relation~\cite{PhysRev.177.2426, Bell:1969ts}
\begin{align}
\partial_\mu \langle J_5^\mu (x) \mathcal{O}(x_1,\ldots,x_n)\rangle = \frac{\alpha_s}{8 \pi}\langle G_{\mu\nu}^a\tilde{G}^a_{\mu\nu}(x)\mathcal{O}(x_1,\ldots,x_n)\rangle
\label{eq:adler}
\end{align}
with $G^a_{\mu\nu}$ the gluon field strength tensor and $\tilde{G}^a_{\mu\nu}=\epsilon^{\mu\nu\rho\sigma} G^a_{\rho \sigma} $ its dual. 

The identity~\eqref{eq:adler} presents a stringent test on scattering amplitudes involving axial vector current insertions, yielding the axial Ward identities.
By replacing $\mathcal{O}(x_1,\ldots,x_n)$ with the correct products of fields for the $ggg$ and $gq\bar{q}$ amplitudes, we can equate the time-ordered product on the left-had side to the matrix element for the three partons coupling to an external current $J^\mu_5$. In momentum space, this becomes simply the contraction $p_4^{\mu} \Omega^{(a)}_\mu$ for the amplitudes under consideration. The right-hand side of~\eqref{eq:adler} can be computed directly as the decay of a pseudoscalar current $A$ to the same three partons with the following interaction Lagrangian~\cite{Kauffman:1993nv}:
\begin{align}
\mathcal{L} = \frac{1}{2}g_A \, \epsilon^{\mu\nu\rho\sigma}G_{\mu\nu}G_{\rho\sigma}A\,,
\end{align}
where $g_A = \alpha_s/(4\pi)$. 
This Lagrangian produces the following Feynman rules for the coupling of the pseudoscalar $A$ to two or three gluons~\cite{Kauffman:1998yg},
\begin{align}
V_2(k^a_{1\mu}, k^b_{2\nu}) &= -i g_A \delta^{ab}\epsilon^{\mu\nu\rho\sigma}k_{1\rho}k_{2\sigma}\,,\\
V_3(k^a_{1\mu}, k^b_{2\nu}, k^c_{1\rho}) &= - g_A \,  g f^{abc}\epsilon^{\mu\nu\rho\sigma}(k_1+k_2+k_3)_{\sigma}\,,
\end{align}
where $g = \sqrt{4\pi \alpha_s}$, while $k_{i}$ are the gluon momenta directed outwards and $a, b, c$ are their
colour indices. The four-gluon vertex vanishes as it is proportional to a
completely antisymmetric combination of $SU(3)$ structure constants.

The IR poles of the pseudoscalar amplitudes are removed as described in section~\ref{sec:UVandIR}, while the renormalisation of $\alpha_s$ and the gluonic 
operator~\cite{Chetyrkin:1998mw,Zoller:2013ixa,Ahmed:2015qpa} combine to give
\begin{align}
    \phi^{(1)} &= S_{\epsilon}^{-1}\overline{\phi}^{(1)} - \frac{3\beta_0}{2\epsilon}\overline{\phi}^{(0)}
    \end{align}
for the removal of UV poles from the bare one-loop amplitude $\overline{\phi}^{(1)}$.

Note that due to the factor $\alpha_s$ in~\eqref{eq:adler}, the tree-level pseudoscalar amplitude relates to the finite remainder of the one-loop axial vector Ward identity, and the one-loop pseudoscalar amplitude to the finite remainder of the two-loop Ward identity. 
The left-hand side of~\eqref{eq:adler} is obtained 
by the contraction $p_4^\mu \Omega_\mu^{(a)}$ and its  
agreement with the pseudoscalar amplitude provides a 
strong check on the correct renormalisation of the 
 axial vector current.

As a consistency  check for the vector 
coupling, we also evaluate $p_4^{\mu} \Omega^{(v)}_\mu$ which has to vanish by the Ward identity.

In order to perform these checks, we first derived an extended tensor basis for the amplitudes in both channels, lifting the unitary gauge condition $p_4 \cdot\epsilon_4=0$. The 6 even and 6 odd tensors in the $gq\bar{q}$ channel are replaced by the following basis of $8+8$ tensors

\begin{align}
A^{v, ext}_{\mu \nu} &= \sum_{i=1}^{6} F_i \,T_i^{E,\mu \nu} + \sum_{i=6}^{8} F_i \,T_{ext,i}^{E,\mu\nu} \nonumber \\
&=  \sum_{i=1}^{6} F_i \,T_i^{E,\mu\nu}  + F_{7} \,\bar{u}(p_2) \slashed{p}_3 u(p_1)p_3^{\mu}p_1^{\nu} + 
F_{8} \,\bar{u}(p_2) \gamma^\nu u(p_1)p_3^{\mu}\,,\label{eqn: ExtendedZgqqV} \\
A^{a, ext}_{\mu\nu} &=\sum_{i=1}^{6} G_i \,T_i^{O,\mu\nu} + \sum_{i=6}^{8} G_i \,T_{ext,i}^{O,\mu\nu}  \nonumber \\
&= \sum_{i=1}^{6} G_i \,T_i^{O,\mu\nu} + G_{7} \,\bar{u}(p_2) \slashed{p}_3 u(p_1)p_3^{\mu}v_A^{\nu} + 
G_{8} \,\bar{u}(p_2) \slashed{v}_A u(p_1)p_3^{\mu}p_1^{\nu}\,,
\label{eqn: ExtendedZgqqA}
\end{align}
where the first 6 structures are always identical to those used for the calculation of the helicity amplitudes.
Contracting the decompositions with $p_4$, we get
{\allowdisplaybreaks
\begin{align}
p_4^{\mu} A^{v, ext}_{ \mu \nu}  &=
\frac{1}{2} \Big\lbrace \big[(s_{12} + s_{13}) F_4 + (s_{12} + s_{23})F_{5} + (s_{13}+s_{23})F_{8}\big]\,  \bar{u}(p_2) \slashed{\gamma}^\nu u(p_1) \nonumber \\ &\phantom{ =\,\,} +
\big[(s_{12} + s_{13}) F_{1} +(s_{12} + s_{23})F_{2} + 2 F_{3} \nonumber \\ &\phantom{ =\,\,}+2 F_{6} + (s_{13}+s_{23})F_{7}\big] \, \bar{u}(p_2) \slashed{p}_3 u(p_1) p_1^{\nu}\Big\rbrace\,,\label{eqn: p4ZgqqV}\\ 
   p_4^\mu A^{a, ext}_{\mu\nu}  &=
 \frac{1}{2} \Big\lbrace \big[(s_{12} + s_{13}) G_{4} + (s_{12} + s_{23})G_{5} + (s_{13}+s_{23})G_{8}\big]\,  \bar{u}(p_2) \slashed{v}_A u(p_1)p_1^\nu \nonumber \\ &\phantom{ =\,\,} +
\big[(s_{12} + s_{13}) G_{1} +(s_{12} + s_{23})G_{2}  +2 G_{6}\nonumber \\ &\phantom{ =\,\,} + (s_{13}+s_{23})G_{7}\big] \, \bar{u}(p_2) \slashed{p}_3 u(p_1) v_A^{\nu} \Big\rbrace\label{eqn: p4ZgqqA}\,.
\end{align}
}

In the $ggg$ channel, lifting the condition $p_4 \cdot \epsilon_4 = 0$ leads to the following basis of $16+16$ tensor structures
\begin{alignat}{3}
A^{v, ext}_{\mu_1 \mu_2 \mu_3 \mu} &= \sum_{i=1}^{12} F_i \,T_i^{E,\mu_1 \mu_2 \mu_3 \mu} &&+ \sum_{i=13}^{16} F_i \,T_{ext,i}^{E,\mu_1 \mu_2 \mu_3 \mu} \nonumber \\
&=  \sum_{i=1}^{12} F_i \,T_i^{E,\mu_1 \mu_2 \mu_3 \mu} &&+ F_{13} \, p_3^{\mu} p_3^{\mu_1} p_1^{\mu_2} p_2^{\mu_3} +
F_{14} \,p_3^{\mu} p_3^{\mu_1} g^{\mu_2\mu_3}  \nonumber \\
& &&+ F_{15}\, p_3^{\mu} p_1^{\mu_2} g^{\mu_1\mu_3} + F_{16} \,p_3^{\mu} p_2^{\mu_3} g^{\mu_1\mu_2}, \\
A^{a, ext}_{\mu_1 \mu_2 \mu_3 \mu} &=\sum_{i=1}^{12} G_i \,T_i^{O,\mu_1 \mu_2 \mu_3 \mu} &&+ \sum_{i=13}^{16} G_i \,T_{ext,i}^{O,\mu_1 \mu_2 \mu_3 \mu}  \nonumber \\
&= \sum_{i=1}^{12} G_i \,T_i^{O,\mu_1 \mu_2 \mu_3 \mu}&&+ G_{13}\,  p_3^{\mu} v_A^{\mu_3} p_3^{\mu_1} p_1^{\mu_2} + 
 G_{14} \, p_3^{\mu} v_A^{\mu_2} p_3^{\mu_1} p_2^{\mu_3} \nonumber \\
& &&+ 
 G_{15} \, p_3^{\mu} v_A^{\mu_1} p_1^{\mu_2} p_2^{\mu_3} + G_{16}\,  p_3^{\mu} \mathcal{T}_O^{\mu_1\mu_2\mu_3}
\end{alignat}
and in this case, the contraction with $p_4$ gives the relations

\begin{align}
p_4^\mu  A^{v, ext}_{\mu_1 \mu_2 \mu_3\mu}  &=
-\frac{1}{2} \Big\lbrace \big[(s_{12} + s_{13}) F_8 + (s_{12} + s_{23})F_9 + 2 F_{12} + (s_{13}+s_{23})F_{15}\big]\, g^{\mu_1 \mu_3} p_1^{\mu_2}  \nonumber \\ &\phantom{ =\,\,} +
\big[(s_{12} + s_{13}) F_{10} +(s_{12} + s_{23})F_{11} + 2 F_{12} + (s_{13}+s_{23})F_{16}\big] \, g^{\mu_1 \mu_2} p_2^{\mu_3}  \nonumber \\ &\phantom{ =\,\,} +\big[(s_{12} + s_{13}) F_{5} + (s_{12} + s_{23})F_{6} + 2 F_{12} + (s_{13}+s_{23})F_{14}\big]\, g^{\mu_2 \mu_3} p_3^{\mu_1}  \nonumber \\ &\phantom{ =\,\,} 
+\big[(s_{12} + s_{13})F_1 + (s_{12} + s_{23})F_2 + 2 F_3 + 
   2 F_4 + 2 F_7 \nonumber \\ &\phantom{ =\,\,}  +  (s_{13}+s_{23})F_{13}\big] p_1^{\mu_3} p_2^{\mu_1}p_3^{\mu_2}  \Big\rbrace\,,\label{eqn: p4ZgggV}\\ 
   p_4^{\mu} A^{a, ext}_{\mu_1 \mu_2 \mu_3\mu}  &=
 -\frac{1}{2} \Big\lbrace  \big[ (s_{12} + s_{13}) F_{11} +(s_{12} + s_{23})F_{12} + (s_{13} + s_{23}) F_{16}\big] \mathcal{T}_O^{\mu_1\mu_2\mu_3} \nonumber \\ &\phantom{ =\,\,}  + \big[ (s_{12} + s_{13}) F_{2} + (s_{12} + s_{23})F_{3} +2 F_6 + 2 F_9 + (s_{13} + s_{23})F_{13} \big] p_3^{\mu_1} p_1^{\mu_2} v_A^{\mu_3} \nonumber \\ &\phantom{ =\,\,} + 
    \big( (s_{12} + s_{13}) F_{4} + (s_{12} + s_{23}))F_{5} +2 F_6 + 2 F_{10} + (s_{13} + s_{23}) F_{14} \big]p_3^{\mu_1} p_2^{\mu_3} v_A^{\mu_2} \nonumber \\ &\phantom{ =\,\,} + 
     \big[(s_{12} + s_{13}) F_{7} + (s_{12} + s_{23})F_{8} + 2 F_9 + 2 F_{10} \nonumber \\ &\phantom{ =\,\,}  + (s_{13} + s_{23}) F_{14} \big]p_1^{\mu_2} p_2^{\mu_3} v_A^{\mu_1}
   \Big\rbrace\label{eqn: p4ZgggA}\,.
\end{align}

For the $ggg$ and $gq\bar{q}$ channel in the axial vector current decay, we used the methods described in section~\ref{sec:tensor} to derive the projectors onto the extended bases, and computed the corresponding form factors. The tensor basis for the pseudoscalar amplitude $Agq\bar{q}$ is given by the 
2 structures on the right hand side of~\eqref{eqn: p4ZgqqA} and for  $Aggg$  by the 
 4 structures in~\eqref{eqn: p4ZgggA}. The associated form factors are obtained with projectors derived as described in section~\ref{sec:tensor}. 
 We can therefore directly compare the pseudoscalar form factors to the linear combinations of 
 axial vector form factors in the Ward identities.

We confirmed that the vector-type Ward identities~\eqref{eqn: p4ZgggV} and~\eqref{eqn: p4ZgqqV} are vanishing at one and two loops. Moreover, we found the expected agreement between tree-level pseudoscalar amplitudes and the one-loop axial vector Ward identities, namely in the $gq\bar{q}$ channel
\begin{align}
p_4^\mu A^{a, ext}_{\mu}  &= +\frac{4 (y-z)}{q^4 y z (1-y-z)} \bar{u}(p_2) \slashed{p}_3 u(p_1) v_A^{\nu} \nonumber\\&
\phantom{ =\,\,}
 -\frac{4 (y+z)}{q^4 y (1-y-z)^2} \bar{u}(p_2) \slashed{v}_A u(p_1)p_1^\nu
\end{align}
and in the $ggg$ channel
\begin{align}
p_4^\mu A^{a, ext}_{\mu_1 \mu_2 \mu_3 \mu}  &=-\frac{4 \left(y^2+y z-y+z^2-z\right)}{3 q^4 y z (1-y-z)}\mathcal{T}_O^{\mu_1\mu_2\mu_3}\nonumber\\&\phantom{ =\,\,}
+\frac{4 \left(y^2-2 y z+2 y+z^2+2 z\right)}{3q^4 y^2 z (1-y-z)}p_3^{\mu_1} p_1^{\mu_2} v_A^{\mu_3}\nonumber\\&\phantom{ =\,\,}
+\frac{4 \left(y^2+4 y z-4 y+4 z^2-4 z+3\right)}{3q^4 y z^2 (1-y-z)}p_3^{\mu_1} p_2^{\mu_3} v_A^{\mu_2}\nonumber \\&\phantom{ =\,\,}
+\frac{4 \left(4 y^2+4 y z-4 y+z^2-4 z+3\right)}{3q^4 y z (1-y-z)^2}p_1^{\mu_2} p_2^{\mu_3} v_A^{\mu_1}\,.
\end{align}
An agreement was also found between the one-loop pseudoscalar amplitudes and two-loop Ward identities, provided the renormalisation of the singlet contributions at two loops includes the finite
renormalisation constant $Z_a^n$ as indicated by the 
last term in~\eqref{eqn: axialrenorm1} and~\eqref{eqn: axialrenorm2}. 

Moreover, the resulting $Vgq\bar{q}$ amplitudes were cross-checked in the decay region with reference~\cite{Gehrmann:2022vuk} up to weight 4. The analytic continuation of the axial vector amplitudes to the production regions, as well as the extension of the form factors to weight 6, are a new result. Note that the top-mass corrections to the pure singlet part of the $Vgq\bar{q}$ amplitudes are vanishing at one loop and are finite at two loops. Therefore, despite the difference in subtraction scheme between this work and~\cite{Gehrmann:2022vuk}, the latter can be straightforwardly combined with our results for the massless contribution to the singlet.

 The $Vggg$ amplitudes coupling to a vector current were checked against~\cite{Gehrmann:2013vga} up to weight 4. The amplitudes with axial vector coupling in the gluonic channel are again a new result and we found agreement for all helicity amplitudes with the numerical result from \code{OpenLoops 2}~\cite{Buccioni:2019sur}.

In the ancillary files, we provide the tree-level amplitudes, and the finite remainders of the one- and two-loop helicity amplitudes continued to the regions in the left columns of tables \ref{tab:crossings} and \ref{tab:crossQ} by specifying the helicity coefficients $\alpha_{i},\ldots,\delta_{i}$. As an essential ingredient for the N$^3$LO $V$+jet amplitudes, we also provide the renormalised helicity amplitude coefficients for the two channels up to $\mathcal{O}(\e^4)$ at one loop and $\mathcal{O}(\e^2)$ at two loops.

\section{Conclusions}\label{sec:conclusions}
We presented the calculation of the two-loop QCD corrections to the production of an electroweak vector boson and a parton in parton-parton annihilation at hadron colliders to higher orders in the dimensional regulator parameter. The calculation relied on
the evaluation of the relevant two-loop Feynman integrals to arbitrary orders in $\epsilon$.
Modern methods for the decomposition of the corresponding amplitudes into independent tensor structures and form factors were essential in computing the missing axial vector contributions to the production of a $Z$ boson and a gluon in gluon fusion.
The evaluation of axial vector contributions in the Larin scheme, while conceptually straightforward with our tensor decomposition, required the integration-by-parts reduction of higher-rank integrals, i.e.\ up to tensor rank 5, compared to the vector contribution which only calls for the reduction of tensor rank 4 integrals. A similar increase by one in the rank had already been observed in similar calculations with axial vector currents.
The presented results are compact, easy to use and have been analytically continued to all regions relevant for $V$+jet production in scattering kinematics. Our amplitudes provide the first ingredient towards the calculation of the finite remainder of three-loop QCD amplitudes for the production of a $V$ boson and a jet at the LHC.

\section*{Acknowledgments}
We are grateful to Federico Buccioni for his help with the validation of our
results against \code{OpenLoops2}. This work was supported in part by the Excellence Cluster ORIGINS funded by the Deutsche Forschungsgemeinschaft (DFG, German Research Foundation) under Germany’s Excellence Strategy – EXC-2094-390783311, by the Swiss National Science Foundation (SNF) under contract 200020-204200, and by the European Research Council (ERC) under the European Union’s research and innovation programme grant agreements 949279 (ERC Starting Grant HighPHun) and 101019620 (ERC Advanced Grant TOPUP).

\bibliographystyle{JHEP}
\bibliography{main}
\end{document}